%% file: main.tex
\documentclass{article}
\usepackage[a4paper, margin=1.3in]{geometry}
\usepackage[utf8]{inputenc}
\usepackage{amsmath}
\usepackage{amsthm}
\usepackage{amssymb}
\usepackage{mathtools}
\usepackage{graphicx}
\usepackage{xcolor}
\usepackage{authblk}

\numberwithin{equation}{section}

\title{The infinite potential well with moving walls}
\author{Kieran Cooney}
\affil{University College Cork}
\date{March 2017}

\begin{document}

\maketitle

\begin{abstract}
In this work the evolution of a wavefunction in an infinite potential well with time dependent boundaries is investigated. Previous methods for wells with walls moving at a constant velocity are summarised. These methods are extended to wells with slowly accelerating walls. The location and time of the revivals in the well of the initial wavefunction are derived using Jacobi's elliptic theta function.
\end{abstract}

\section{Introduction}

The infinite potential well (which we will sometimes just refer to as a well) is one of the earliest examples most students encounter when first studying quantum mechanics. Mathematically, this problem may be phrased as finding a solution to the time dependent Schrodinger equation:
\begin{equation}
    i\hbar\frac{\partial \psi}{\partial t}=-\frac{\hbar^{2}}{2m}\frac{\partial^{2}\psi}{\partial x^{2}}
\end{equation}
subject to the boundary conditions $\psi(x=0,t)=\psi(x=w,0)=0$. The time dependence may be removed from the problem by instead considering the energy eigenmodes $\psi(x,t)=\psi_{n}(x)\exp\left(-iE_{n}/\hbar\right)$, which yields the time independent Schrodinger equation:
\begin{equation}
\label{Time independent Schrodinger equation}
E_{n}\psi_{n}(x)=-\frac{\hbar^{2}}{2m}\frac{d^{2}\psi_{n}(x)}{d x^{2}}
\end{equation}
The solutions to \eqref{Time independent Schrodinger equation} are
\begin{equation}
\psi_{n}(x)=\sqrt{\frac{2}{w}}\sin\left(\frac{n\pi x}{w}\right)
\end{equation}
for $n$ a positive integer, and $E_{n}=(\hbar n \pi)^{2}/2mw^{2}$.
There solutions to equation \eqref{Time independent Schrodinger equation} are in bijection with the positive integers, which leads naturally to energy quantisation. 

Considering the simplicity of the above example, a logical extension is to consider a wavefunction $\psi(x,t)$ evolving in an infinite potential well with walls moving along arbitrary paths $w_{1}(t)$ and $w_{2}(t)$. This is the problem we are concerned with in this work. The flavour of the text is towards exposition; the results may be presented much more concisely than they are here. The intention is to make the results as approachable as possible to the student curious about extending their simple example.

The earliest mention of the well with moving walls appears to be in 1953 by Hill and Wheeler \cite{HillWheeler}. Their paper is a detailed theoretical study of nuclear physics, and any mention of particles in boxes is hidden in an appendix. In 1969, Doescher and Rice provide a full set of solutions for an infinite potential well with one wall moving at a fixed velocity \cite{Doescher}. Much work was done on wavefunctions subjected to time dependent boundary conditions throughout the 80's and early 90's \cite{Berry:1984,LevyLeBlond,Greenberger,Pinder,Makowski:91,Pereshogin,Makowski:92,Dodonov,MORALES}. More recently, the infinite potential well with moving walls has been useful in the field of shortcuts to adiabaticity \cite{Cervero,Campo}. In section \ref{Linear wells}, we will review some of the different methods that have been applied to the well with walls moving at a constant velocity. In section \ref{Nonlinear box and revivals} we will extend these methods to wells whose walls undergo a small acceleration.

Although a simple model, the infinite potential well yields some very interesting dynamics. wavefunctions in a well experience \emph{revivals}; after evolving for a certain time the wavefunction will appear as a sum of copies of the original \cite{Berry:1996,AronsteinStroud,Berry:2001}. In sections \ref{Linear revivals} and \ref{Nonlinear box and revivals} we will derive the revivals for a wavefunction in a well with walls moving at a constant velocity or with a small acceleration, respectively. 

In the well, $\psi(x,t)$ just obeys the free space Schrodinger equation subject to some boundary conditions. In some instances, it will be easier to dismiss the boundary conditions and consider a wavefunction evolving in free space instead. In section \ref{Linear wells} we will examine those coordinate transformations which preserve the free space Schrodinger equation \cite{Niederer}. In section \ref{Nonlinear box and revivals} we will extend these transformations to non-inertial reference frames and derive the resulting non-inertial forces \cite{Rosen,Takagi:1}.

\section{The infinite potential well with walls moving at a fixed velocity}
\label{Linear wells}

\begin{figure}[h]
    \begin{center}
    \def\svgwidth{0.8\columnwidth}
    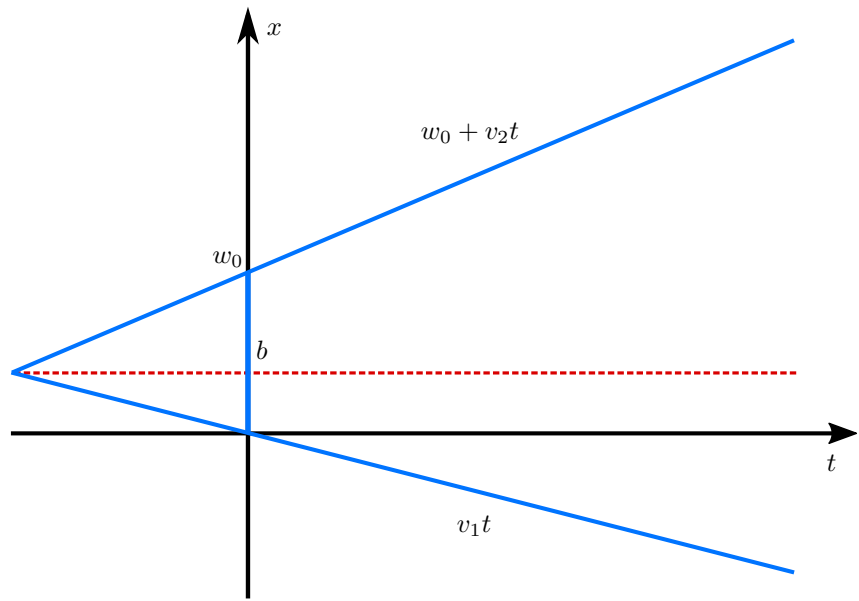
    \caption{An infinite potential well with walls moving at a constant velocity, intersecting at the point $b$.}
    \label{fig:1}
    \end{center}
\end{figure}

Consider a particle with mass $m$ that is constrained between the two walls $w_{2}(t)=w_{0}+v_{2}t$ and $w_{1}(t)=v_{1}t$, as in figure \ref{fig:1}. The instantaneous width of the well is then $w(t)=w_{0}+(v_{2}-v_{1})t$ and the rate of increase of this width is $\Delta v=v_{2}-v_{1}$. Then the wavefunction $\psi(x,t)$ must satisfy Schrodinger's equation

\begin{equation}
    \label{Schrodinger's equation}
    i\hbar\frac{\partial \psi}{\partial t}=-\frac{\hbar^{2}}{2m}\frac{\partial^{2}\psi}{\partial x^{2}}
\end{equation}
subject to the boundary conditions $\psi(w_{2}(t),t)=\psi(w_{1}(t),t)=0$. The following is a set of solutions for this problem:
\begin{equation}
\label{The solution set}
    \psi_{n}(x,t)=\sqrt{\frac{2}{w(t)}}\sin\left(\frac{n\pi(x-w_{1}(t))}{w(t)}\right)\exp\left(i\theta(x,t)-i\frac{mE_{n}^{0}w_{0}t}{\hbar w(t)}\right)
\end{equation}
where $E_{n}^{0}=(n^{2}\hbar^{2}\pi^{2})/2mw_{0}^{2}$ is the energy eigenvalue for the $n$-th eigenstate infinite potential well with constant width $w_{0}$ and
\begin{equation}
\label{theta definition}
\theta(x,t)
=\dfrac{m\left(\Delta v x^{2}+2v_{1}w_{0}x-v_{1}^{2}w_{0}t\right)}{2\hbar w(t)}
\end{equation}
Note that $\theta(x,t)$ is independent of the mode number $n$.

The solutions \eqref{The solution set} are very similar to the adiabatic solutions $\Psi_{n}(x,t)$:
\begin{equation}
\label{Adiabatic solutions}
\Psi_{n}(x,t)=\sqrt{\frac{2}{w(t)}}\sin\left(\frac{n\pi(x-w_{1}(t))}{w(t)}\right)
\exp\left(i\left(-\int_{0}^{t}\frac{E_{n}(t')}{\hbar}\,dt'+\gamma_{n}(t)\right)\right)
\end{equation}
Here $\sqrt{2/w(t)}\sin\left(n\pi(x-w_{1}(t))/w(t)\right)$ is the instantaneous eigenmode, $E_{n}(t)$ is the instantaneous eigenvalue and $\gamma_{n}(t)$ is the geometric-Berry phase \cite{BerryPhase}. 
The shape of the two solutions are the same, that being the instantaneous eigenmodes. Thus, the solutions \eqref{The solution set} differ from the solutions \eqref{Adiabatic solutions} by a phase term.
The time evolution of each $\psi_{n}(x,t)$ is governed by the phase term $\exp\left(-iE_{n}^{0}w_{0}t/\hbar w(t)\right)$, which is exactly the dynamic phase term of the adiabatic approximation. Let $E_{n}(t)=(n^{2}\hbar^{2}\pi^{2})/2mw(t)^{2}$, the energy eigenvalue of the $n$th instantaneous eigenmode at time $t$. Then 
\begin{equation}
\exp\left(-i\frac{E_{n}^{0}w_{0}t}{\hbar w(t)}\right)=\exp\left(-i\int_{t'=0}^{t}\frac{E_{n}(t')}{\hbar}\,dt'\right)
\end{equation}
This may be verified by direct integration.
\begin{equation}
\begin{aligned}
\int_{t'=0}^{t}\frac{E_{n}(t')}{\hbar}\,dt'&=\int_{t'=0}^{t}\frac{n^{2}\hbar\pi^{2}}{2m(w_{0}+\Delta v t')^{2}}\,dt' \\
&=\frac{n^{2}\hbar\pi^{2}}{2m\Delta v}\left.\left(\frac{1}{w_{0}+\Delta v t'}\right)\right|_{0}^{t} \\
&=-\frac{n^{2}\hbar\pi^{2}}{2m\Delta v}\left(\frac{1}{w_{0}+\Delta v t}-\frac{1}{w_{0}}\right) \\
&=-\frac{n^{2}\hbar\pi^{2}}{2m\Delta v}\left(\frac{1}{w_{0}+\Delta v t}-\frac{1}{w_{0}}\right) \\
&=\frac{n^{2}\hbar\pi^{2}}{2mw(t)w_{0}} = \frac{E_{n}^{0}w_{0}t}{\hbar w(t)}
\end{aligned}
\end{equation}
The geometric phase may be calculated by
\begin{equation}
\label{Berry equation}
\gamma_{n}(t)=i\int_{R(0)}^{R(t)}\left<n(R(t'))\left|\nabla_{R}\right|n(R(t'))\right>dR'
\end{equation}
where $R(t)=\left(w_{1}(t),w_{2}(t)\right)$ and $\left|n(R(t))\right>$ is the $n$th instantaneous eigenmode with lower and upper walls $w_{1}(t)$ and $w_{2}(t)$ respectively:
\begin{equation}
\left|n(R)\right>=\sqrt{\frac{2}{(w_{2}-w_{1})}}\sin\left(\frac{n\pi(x-w_{1})}{w_{2}-w_{1}}\right)
\end{equation}
For the sake of abbreviation, let $u=n\pi(x-w_{1})/(w_{1}-w_{1})$ and $w=w_{2}-w_{1}$. Then
\begin{equation}
\begin{aligned}
\left<n(R)\left|\frac{\partial}{\partial w_{1}}\right|n(R)\right>&= \frac{-2}{w}\int_{w_{1}}^{w_{2}}\sin(u)\left(\frac{\sin(u)}{2w}+\frac{u}{w}\cos(u)\right) \,dx \\
&=\frac{-1}{w}\int_{0}^{n\pi}\sin^{2}(u)+2u\sin(u)\cos(u)\,du
\end{aligned}
\end{equation}
This trigonometric integral may be evaluated directly, or by observing that $d\left(u \sin^{2}(u)\right)/du$ is exactly the integrand. Either way, the integral vanishes. By symmetry, $\left<n(R)\left|\partial/\partial w_{2}\right|n(R)\right>$ must vanish also. Thus, $\left<n(R(t'))\left|\nabla_{R}\right|n(R(t'))\right>=0$ and the geometric phase $\gamma_{n}(t)$ is trivial.
Thus $\psi_{n}(x,t)$ is very close to the adiabatic solution $\Psi_{n}(x,t)$ except for the extra phase term $\theta(x,t)$. In the following sections we will try to explain some of the physical nature of this extra factor.

$\theta(x,t)$ simplifies if we specify whether $v_{1} = v_{2}$ or $v_{1}\neq v_{2}$. If $v_{1}=v_{2}=v$, $\Delta v =0$ and $w_{t}=w$ is constant. Then $\theta(x,t)$ simplifies to

\begin{equation}
\label{The parallel solution set}
\theta(x,t)=i\frac{mvx-(mv^{2}t)/2}{\hbar}
\end{equation}

On the other hand, suppose that $v_{1}\neq v_{2}$. Then $\Delta v \neq 0$, and $1/\Delta v$ is finite. Adding a constant phase of  $(mv_{1}^{2}w_{0})/(2\hbar\Delta v)$ to $\psi_{n}(x,t)$ allows us to complete the square in $x$:
\begin{gather}
\label{The non parallel solution set calculation}
\begin{aligned}
\theta(x,t)&=
\frac{m\left(\Delta v x^{2}+2v_{1}w_{0}x-v_{1}^{2}w_{0}t\right)}{2\hbar w(t)} +\frac{mv_{1}^{2}w_{0}}{2\hbar\Delta v}\\
&=\frac{m\Delta v}{2\hbar w(t)}\left(x^{2}+\frac{2v_{1}w_{0}}{\Delta v}x+\frac{v_{1}^{2}w_{0}(w(t)-\Delta v t)}{\Delta v^{2}}\right) \\
&=\frac{m\Delta v}{2\hbar w(t)}\left(x^{2}+\frac{2v_{1}w_{0}}{\Delta v}x+\frac{v_{1}^{2}w_{0}^{2}}{\Delta v^{2}}\right) \\
\end{aligned} \\
\label{The non parallel solution set}
\Rightarrow \theta(x,t)=\frac{m\Delta v}{2\hbar w(t)}(x-b)^{2}
\end{gather}
where 
\begin{equation}
\label{b definition}
b=-\frac{v_{1}w_{0}}{\Delta v}
\end{equation}
is the point of intersection of the two walls, as may be verified by simple geometry. 

If $w_{0}=0$, $w(t)=\Delta v t$ and $\theta(x,t)$ further simplifies to $mx^{2}/2\hbar t$. However in this case, the solutions $\psi_{n}(x,t)$ become singular as $t\to0$. This is closely related to the Appell transformation \eqref{Appell transformation}.

Note that if we take the limit $\Delta v\rightarrow 0$, the walls become parallel and $b\rightarrow \infty$. Equation \eqref{The parallel solution set} may thus be derived from equation \eqref{The non parallel solution set} in this limit. The calculation of this limit closely follows equation \eqref{The non parallel solution set calculation}. In section \ref{DR section}, equation \eqref{The parallel solution set} will be derived from a simple application of moving reference frames. Thus as the non parallel case is more general and less trivial than the parallel case, it is often enough to consider equation \eqref{The non parallel solution set} in place of equation \eqref{theta definition}.

There are a number of ways to arrive at the solution set \eqref{The solution set}, although they are all similar. We will now present some of these methods in the context of some other author's studies.

\subsection{Doescher and Rice's method}
\label{DR section}

In 1969, Doescher and Rice \cite{Doescher} presented the following set of solutions to the infinite potential well with one moving wall, i.e. solutions to Schrodinger's equation \eqref{Schrodinger's equation} such that $\psi(x,t)=0$ for $x=0$ and $x=w_{0}+\Delta vt$.
\begin{equation}
\label{Doescher Rice solutions}
\psi_{n}(x,t)=\sqrt{\frac{2}{w(t)}}\sin\left(\frac{n\pi x}{w(t)}\right)\exp\left(i\dfrac{m\Delta vx^{2}-2E_{n}^{0}w_{0}t}{2\hbar w(t)}\right)
\end{equation}
No derivation was provided for these solutions, just their existence. The most straightforward way to verify that these satisfy Schrodinger's equation is to first show that
\begin{equation}
\label{DR phase solutions}
\varepsilon_{n}(x,t)=\sqrt{\frac{2}{w(t)}}\exp\left(i\dfrac{2\hbar n\pi x+m\Delta vx^{2}-2E_{n}^{0}w_{0}t}{2\hbar w(t)}\right)
\end{equation}
satisfies Schrodinger's equation, and then show that $\psi_{n}(x,t)=(\varepsilon_{n}(x,t)-\varepsilon_{\texttt{-}n}(x,t))/2i$ satisfies the appropriate boundary conditions. To show that $\varepsilon_{n}(x,t)$ is a solution for $n \in \mathbb{N}$, we write
\begin{equation}
\label{Epsilon exponential substitution}
\varepsilon _{n}(x,t)=\sqrt{\frac{2}{w(t)}}\exp\left(if(x,t)\right)
\end{equation}
where $f(x,t)$ is a real valued function. Substituting equation \eqref{Epsilon exponential substitution} into equation \eqref{Schrodinger's equation} yields
\begin{equation}
\label{Exponential Schrodinger's equation 1}
i\hbar\left(if_{t}-\frac{\Delta v}{2w(t)}\right)\varepsilon=-\frac{\hbar^{2}}{2m}\left(if_{xx}-f_{x}^{2}\right)\varepsilon
\end{equation}
Comparing real and imaginary parts of equation \eqref{Exponential Schrodinger's equation 1} gives two coupled partial differential equations:
\vspace{-3mm}
\begin{gather}
\label{Exponential Schrodinger's equation 2}
-\hbar f_{t} = \frac{\hbar^{2}}{2m}f_{x}^{2} \\
\label{Exponential Schrodinger's equation 3}
\frac{\hbar \Delta v}{2w(t)}=\frac{\hbar^{2}f_{xx}}{2m}
\end{gather}
It is straightforward to check that $f(x,t)$, as defined in equation \eqref{DR phase solutions},  satisfies these equations. 

We may obtain the solution for two moving walls by using a moving reference frame. For the free space Schrodinger equation \eqref{Schrodinger's equation}, if $\psi(x,t)$ is some solution then 
\begin{equation}
\label{moving reference frame transformation}
\psi(x+vt,t)\exp\left(\frac{i}{\hbar}\left(-mvx-\frac{imv^{2}t}{2}\right)\right)
\end{equation}
is the boosted solution in a moving reference frame. To see this, consider a plane wave solution to the free space Schrodinger equation, $\exp\left(i\left(px-p^{2}t/2m\right)/\hbar\right)$. Now ``translate'' the momentum $p\mapsto p-mv$;
\begin{multline}
\label{Plane wave boost}
    \exp\left(\frac{i}{\hbar}\left((p-mv)x-\frac{(p-mv)^{2}}{2m}t\right)\right) \\
    \begin{aligned}
     &=\exp\left(\frac{i}{\hbar}\left(px-mvx-\frac{p^{2}}{2m}t+pvt-\frac{mv^{2}}{2}t\right)\right) \\
     &=\exp\left(\frac{i}{\hbar}\left(p(x+vt)-\frac{p^{2}}{2m}t\right)\right)\exp\left(\frac{i}{\hbar}\left(-mvx-\frac{mv^{2}}{2}t\right)\right)
    \end{aligned}
\end{multline}
Equation \eqref{Plane wave boost} is also a solution to the free space Schrodinger equation, and has the form of a plane wave in a moving reference frame $\exp\left(i\left(p(x+vt)-(p^{2}/2m)t\right)/\hbar\right)$ multiplied by a phase term $\exp\left(-i(mvx+mv^{2}t/2)/\hbar\right)$. However, the phase term $\exp\left(-i(mvx+mv^{2}t/2)/\hbar\right)$ is independent of the initial momentum $p$, and so equation \eqref{moving reference frame transformation} is also a solution to the free space Schrodinger equation.
Note that by moving a well with fixed walls to a moving reference frame and applying \eqref{moving reference frame transformation}, equation \eqref{The parallel solution set} is derived.

\begin{figure}[h]
    \def\svgwidth{\columnwidth}
    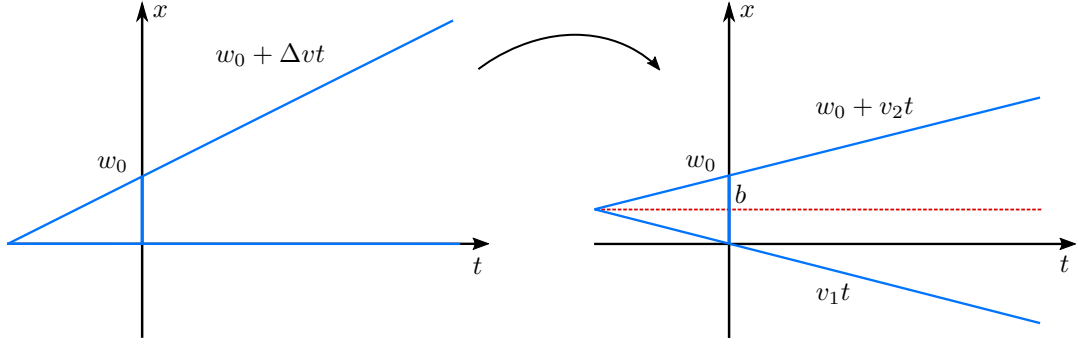
    \caption{A mapping of a well with one moving wall to a well with two moving walls using a moving reference frame.}
    \label{fig:boosted reference frame}
    \centering
\end{figure}

Thus, consider an infinite potential well with $w_{1}(t)=0$ and $w_{2}(t)=w_{0}+(v_{2}-v_{1})t$. The relevant Doescher-Rice solutions are 
\begin{equation}
\psi_{n}(x,t)=\sqrt{\frac{2}{w(t)}}\sin\left(\frac{n\pi x}{w(t)}\right)\exp\left(i\dfrac{m(v_{2}-v_{1})x^{2}-2E_{n}^{0}w_{0}t}{2\hbar w(t)}\right)
\end{equation}
By switching reference frames $(x,t)\mapsto(x-v_{1}t,t)$ as in figure \ref{fig:boosted reference frame}, equation \eqref{moving reference frame transformation} implies that the following expression is a solution in a well with walls $w_{1}(t)=v_{1}t$ and $w_{2}(t)=w_{0}+v_{2} t$:
\begin{equation}
\label{Boosted Doescher Rice}
\sqrt{\frac{2}{w(t)}}\sin\left(\frac{n\pi (x-v_{1}t)}{w(t)}\right)\exp\left(\frac{i}{\hbar}\left(\frac{m\Delta v (x-v_{1}t)^{2}-2E_{n}^{0}w_{0}t}{2w(t)}+mv_{1}x-\frac{mv_{1}^{2}t}{2}\right)\right)
\end{equation}
We may simplify the phase of the above solution:
\begin{multline}
\label{Boosted Doescher Rice algebra}
\frac{m}{2\hbar w(t)}\left(\Delta v(x-v_{1}t)^{2}+2v_{1}w(t)x-v_{1}^{2}w(t)t\right) \\
\begin{aligned}
 &=\frac{m}{2\hbar w(t)}\left(\Delta v(x^{2}-2xv_{1}t+v_{1}^{2}t^{2})+2v_{1}(w_{0}+\Delta v\,t)x-v_{1}^{2}(w_{0}+\Delta v\,t)t\right) \\
 &=\frac{m}{2\hbar w(t)}\left(\Delta vx^{2}+2v_{1}w_{0}x-v_{1}^{2}w_{0}t\right) = \theta(x,t)
\end{aligned}
\end{multline}
and thus equation \eqref{Boosted Doescher Rice} agrees with equation \eqref{The solution set}. Equation \eqref{Boosted Doescher Rice} (or the first line of \eqref{Boosted Doescher Rice algebra}) is a useful variation of \eqref{The solution set} as the $x$ component of $\theta$ is independent on the initial width $w_{0}$.

\subsection{Greenberger's method}
\label{GB section}

In 1988, Greenberger also arrived at the solution \eqref{The solution set} using a transformation of variables. Once again, consider the free space Schrodinger equation \eqref{Schrodinger's equation}. To create a stretching in the spatial dimension, Greenberger introduced the following new variables
\begin{equation}
y=\frac{x}{w(t)} \qquad t'=t
\end{equation}
where $w(t)=w_{0}+\Delta v t$. Then applying the chain rule for partial derivatives:
\begin{equation}
\begin{gathered}
\frac{\partial \psi}{\partial x} = \frac{\partial \psi}{\partial y}\frac{\partial y}{\partial x} + \frac{\partial \psi}{\partial t'}\frac{\partial t'}{\partial x} = \frac{1}{w}\frac{\partial \psi}{\partial y} \\
\frac{\partial \psi}{\partial t} = \frac{\partial \psi}{\partial y}\frac{\partial y}{\partial t} + \frac{\partial \psi}{\partial t'}\frac{\partial t'}{\partial t} =  -\frac{\partial \psi}{\partial y}\frac{y\Delta v}{w(t)} + \frac{\partial \psi}{\partial t'}
\end{gathered}
\end{equation}
In these stretched coordinates, Schrodinger's equation becomes \eqref{Schrodinger's equation} becomes
\begin{equation}
\label{Greenberger 0}
\frac{-\hbar^{2}}{2mw^{2}}\frac{\partial^{2}\psi}{\partial y^{2}}+i\hbar\frac{\Delta v}{w(t)}y\frac{\partial \psi}{\partial y}=i\hbar \frac{\partial \psi}{\partial t}
\end{equation}
We would like to transform $\psi$ somehow so as to compensate for this stretching, and thus simplify equation \eqref{Greenberger 0}. One way to proceed is to introduce a kind of  integrating factor so as to cancel the $\partial \psi /\partial y$ term. Thus let $\psi(y,t)=\chi(y,t)f(y,t)$, where $f$ is an integrating factor. Then Schrodinger's equation for $\chi(y,t)$ is
\begin{equation}
\label{Greenberger 1}
\frac{-\hbar^{2}}{2mw^{2}}\left(f_{yy}\phi+2f_{y}\chi_{y}+f\chi_{yy}\right)+i\hbar\frac{\Delta v}{w(t)}y\left(f_{y} \chi + f \chi_{y} \right) = i \hbar\left(f\chi_{t}+f_{t}\chi\right)
\end{equation}
By imposing that the coefficient of $\chi_{y}$ should vanish, we arrive at the following differential equation for $f(y,t)$:
\begin{equation}
\begin{gathered}
-\frac{\hbar^{2}}{mw^{2}}f_{y}+i\frac{\hbar \Delta v}{w(t)}yf=0 \\
\Rightarrow \frac{f_{y}}{f} = \ln (f)' = \frac{im\Delta v \,w }{\hbar}y \\
\Rightarrow f(y,t)=\exp\left(i\frac{m \Delta v \,w(t)y^{2}}{2\hbar}\right)
\end{gathered}
\end{equation}
Thus 
\begin{equation}
\chi(y,t)=\psi(y,t)\exp\left(-i\frac{ m \Delta v \,w(t)y^{2}}{2\hbar} \right)
\end{equation} 
Upon substitution into equation \eqref{Greenberger 1}, we find that $\chi(y,t)$ satisfies a variant of the free space Schrodinger equation \eqref{Schrodinger's equation}:
\begin{equation}
\label{Greenberger 2}
\frac{-\hbar^{2}}{2m}\frac{\partial ^{2}\chi}{\partial y^{2}} = i\hbar \left(w(t)^{2}\frac{\partial \chi}{\partial t}+\frac{\Delta v\,w}{2}\chi\right)
\end{equation}
To remove the $\chi$ term from \eqref{Greenberger 2}, we simply introduce a normalisation term. Define $\phi(y,t)$ as
\begin{equation}
\label{Greenberger 2.5}
\phi(y,t)=\sqrt{w(t)}\chi(y,t)=\sqrt{w(t)}\psi(y,t)\exp\left(-i\frac{ m \Delta v \,w(t)y^{2}}{2\hbar} \right)
\end{equation}
$\phi(y,t)$ has the advantage over $\chi(y,t)$ that if $\psi(x,t)$ is normalised in $x$ for a given value of $t$, then $\phi(y,t)$ is normalised in $y$ for the same value of $t$. Substituting $\phi$ for $\chi$, equation \eqref{Greenberger 2} then becomes 
\begin{equation}
\label{Greenberger 3}
\frac{-\hbar^{2}}{2m}\frac{\partial ^{2}\phi}{\partial y^{2}}=i\hbar\frac{\partial \phi}{\partial t}w^{2}(t)
\end{equation}
To remove the $w(t)^{2}$ term, we need to re-parametrise time. Denote this re-parametrisation by $\tau(t)$. Then by the chain rule:
\begin{equation}
    \frac{\partial \phi}{\partial t}=\frac{\partial \phi}{\partial \tau}\frac{d\tau}{dt}
\end{equation}
Setting $d\tau/dt=w^{-2}(t)$ and solving for $\tau$:
\begin{gather}
\label{tau(t)}
\tau(t)=\int_{0}^{t} w^{-2}(t')\,dt'=\frac{1}{\Delta v}\left(\frac{1}{w_{0}}-\frac{1}{w(t)}\right)=\frac{t}{w_{0}w(t)}
\\
\label{t(tau)}
\Leftrightarrow t(\tau) = \frac{w_{0}^{2}\tau}{1-w_{0}\Delta v\tau}
\end{gather} and
\begin{equation}
\frac{-\hbar^{2}}{2m}\frac{\partial ^{2}\phi}{\partial y^{2}}=i\hbar\frac{\partial \phi}{\partial \tau}
\end{equation}
which is of course just the free space Schrodinger equation \eqref{Schrodinger's equation}. This allows us to generate solutions in the stretched coordinates from solutions in rectangular coordinates. By solving for $\psi(x,t)$ in equation \eqref{Greenberger 2.5}, we see that if $\phi(y,\tau)$ is a free space solution then
\begin{equation}
\label{Greenberger 4}
\psi(x,t)=\frac{1}{\sqrt{w(t)}}\phi\left(\frac{x}{w(t)},t\right)\exp\left(i\frac{m\Delta v\,x^{2}}{2\hbar w(t)}\right)
\end{equation}
is also a free space solution, where $t(\tau)$ is defined above.

\begin{figure}[h]
    \def\svgwidth{\columnwidth}
        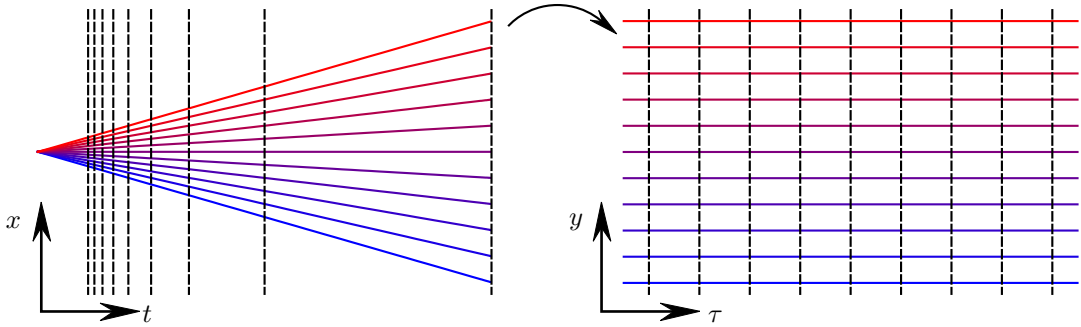
    \caption{Under the transformation $(x,t)\mapsto(y,\tau)$, all the lines in the $(x,t)$ plane passing through a particular point are mapped to horizontal lines in the $(y,\tau)$ plane.}
    \label{fig:Greenberger transformation}
    \centering
\end{figure}

Equation \eqref{Greenberger 4} is a useful technique for generating solutions to the free space Schrodinger equation. In particular, it allows us to move between the two coordinate systems in figure \ref{fig:Greenberger transformation}. The vertical dashed lines in the $(x,t)$ plane are mapped to those in the $(y,\tau)$ plane. Note that the gaps between the lines increase dramatically in the $(x,t)$ plane, even though they are fixed in the $(y,\tau)$ plane. This is because we have defined $\tau$ so that $\Delta t = w^{2}(t)\Delta \tau$. As $\phi(y,\tau)$ evolves as a free space wavefunction, in some sense we expect that the time evolution of  $\psi(x,t)$ should be slower for larger $w(t)$.

As a simple example, consider the trivial solution to equation \eqref{Greenberger 3} $\phi(y,t)=1$. (For the reader not comfortable using a wavefunction that is not normalisable, this argument works for very wide wave packets also.) Then applying equation \eqref{Greenberger 4}, we see that
\begin{equation}
\psi(x,t)= \frac{1}{\sqrt{w(t)}}\exp\left(i\frac{m\Delta v x^{2}}{2\hbar w(t)}\right)
\end{equation}
is a solution to the free space solution. This may easily be checked directly. The momentum of this wavefunction at a given $(x,t)$ may be calculated by applying the momentum operator $\hat{p}=-i\hbar \partial_{x}$:
\begin{equation}
\begin{aligned}
\hat{p}\psi(x,t)&=-i\hbar\frac{\partial}{\partial x}\frac{1}{\sqrt{w(t)}}\exp\left(i\frac{m\Delta v x^{2}}{2\hbar w(t)}\right) \\
&= m \Delta v \frac{x}{w(t)} \psi
\end{aligned}
\end{equation}
This is as expected; the momentum of the wavefunction increases linearly with the displacement from the $x=0$ axis. Takagi has likened this to Hubble's law of cosmology \cite{Takagi:1}. The constant of proportionality is such that when $x=w(t)$, the momentum is that of a classical particle with mass $m$ moving at the same speed as the coordinate line. Thus the ``geometric" phase term exists to keep track of the momentum, as noted by Greenberger. 

By the above, we expect that iterating the Greenberger transformation should generate an infinite family of solutions that satisfy Schrodinger's equation. In fact, Greenberger transformations are closed in the sense that the product of two Greenberger transformations is another Greenberger transformation. This idea is formalised using Niederer's transformations in appendix \ref{Niederer appendix}.

Returning to the problem of the infinite potential well with moving walls, consider an infinite potential well with walls $y=0,1$ in the $(y,\tau)$ plane. Then the eigenmodes are
\begin{equation}
\phi(y,\tau)=\sin(n\pi y)\exp\left(-i\frac{n^{2}\hbar\pi^{2}}{2m}\right)
\end{equation}
Applying Greenberger's transformation \eqref{Greenberger 4} to these eigenmodes exactly reproduces the Doescher Rice solutions \eqref{Doescher Rice solutions} for one moving wall. 

We may now deduce the more general solutions \eqref{The solution set} in a number of ways. Firstly, we may proceed as in section \ref{DR section} and apply a Galilean boost to the solutions for one moving wall. Secondly, we may use Niederer's transformation \eqref{Niederer transformation} from appendix \ref{Niederer appendix} and deduce a transformation $g$ which sends two fixed walls to two non-accelerating moving walls. This would also necessitate calculating $f_{g}$. 

Thirdly, note that in figure \ref{fig:Greenberger transformation} there is a translational symmetry in the $(y,\tau)$ plane along the $y$ direction.  However translation in $y$ corresponds to switching between the coincident lines in the $(x,t)$ plane, provided $\Delta v \neq 0$. Let $y'=y+d$, and $x'$ the result of this translation in the $(x,t)$ plane. Then
\begin{equation}
x'=(y+d)w(t)=yw(t)+dw_{0}+d\Delta v t = x+dw_{0}+d\Delta v t
\end{equation}
By setting $d=-v_{1}/\Delta v$:
\begin{equation}
x'=x-v_{1}(t)+b
\end{equation}
which is just a moving reference frame. $b$ is defined as in equation \eqref{b definition}.
Thus equation \eqref{Greenberger 4} may be generalised; if $\phi(y,\tau)$ is a solution to the free space Schrodinger equation, then so is
\begin{equation}
\frac{1}{\sqrt{w(t)}}\phi\left(\frac{x-v_{1}t+b}{w(t)},t\right)\exp\left(i\frac{m\Delta v\,x^{2}}{2\hbar w(t)}\right)
\end{equation}
Translating $x\mapsto x-b$ then shows that
\begin{equation}
\label{Inverse Greenberger transformation}
\psi(x,t)=\frac{1}{\sqrt{w(t)}}\phi\left(\frac{x-v_{1}t}{w(t)},t\right)\exp\left(i\frac{m\Delta v\,(x-b)^{2}}{2\hbar w(t)}\right)
\end{equation}
is also a solution to the free space Schrodinger equation. 

With this in mind, for a given infinite potential well with moving walls with wavefunction $\psi(x,t)$, define the \emph{Greenberger transformation} $G(\psi(x,t))=\phi(y,\tau)$ by
\begin{equation}
\label{Greenberger transformation}
\phi(y,\tau)=\sqrt{w(t)}\psi(y,\tau)\exp(-i\theta(y,\tau))
\end{equation}
where
\begin{gather}
\label{theta y definition}
\theta(y,t)=\frac{m\Delta v w(t)}{2\hbar}(y-c)^{2} \\
y=(x-v_{1}t)/w(t) \qquad \tau=\int_{0}^{t}\frac{dt'}{w(t')^{2}} \qquad c= (b-v_{1}t)/w(t) = \frac{-v_{1}}{\Delta v}
\end{gather}
Clearly $\phi(y,\tau)$ satisfies Schrodinger's equation and the boundary conditions $\phi(0,\tau)=\phi(1,0)$, and so belongs in an infinite potential well with fixed width. Similarly for any such $\phi(y,\tau)$, define the inverse Greenberger transformation $\psi(x,t)=G^{-1}(\phi(y,\tau))$ by equation \eqref{Inverse Greenberger transformation}. $\psi(x,t)$ is a solution to the infinite potential well with moving walls. Greenberger's transformation shows that the solutions to both problems are in bijection with each other. In particular, $G^{-1}\left(\sqrt{2}\sin(n\pi y)\exp(-in^{2}\pi^{2}\hbar \tau/2m)\right)$ yields the solutions $\psi_{n}(x,t)$ \eqref{The solution set} for $\Delta v \neq 0$. 

\subsection{Berry and Klein's method}
\label{BK method}

In 1984, Berry and Klein \cite{Berry:1984} studied the dynamics of a wavefunction $\psi(x,t)$ subject to an expanding potential $\alpha(l(t))V(x/l(t))$, where $l(t)$ is the linear scale factor and $\alpha(l)$ scales the strength of the expanding potential. Their analysis was for 3D systems, but again we restrict to the 1D case. Thus $\psi(x,t)$ must be a solution to
\begin{equation}
\label{Berry 1}
    i\hbar\frac{\partial \psi}{\partial t} = \frac{-\hbar^{2}}{2m}\frac{\partial ^{2}\psi}{\partial x^{2}} + \alpha(l(t))V\left(\frac{x}{l(t)}\right)\psi
\end{equation}
The normalised variables $\rho = x/l(t)$ and $\tau=\int_{0}^{t}dt/l(t)^{2}$ are introduced, so that 
\begin{equation}
\label{Berry 2}
    i\hbar\frac{\partial \psi}{\partial \tau} = \frac{-\hbar^{2}}{2m}\frac{\partial ^{2}\psi}{\partial \rho^{2}} + \left(V(\rho)+\frac{1}{2}k\rho^{2}\right)\psi
\end{equation}
where
\begin{equation}
\label{Berry 3}
\psi(x,t)=\frac{\phi(\rho,\tau)}{\sqrt{l}}\exp\left(i\frac{mll'\rho^{2}}{2\hbar}\right)
\end{equation}  
and
\begin{equation}
\label{Berry 3.5}
l(t)=\sqrt{at^{2}+2bt+c} \qquad \qquad
ac-b^{2}=k/m \qquad \qquad
\end{equation}
We will not go through the details of the above coordinate transformation here as it is very similar to the procedure carried out in section \ref{GB section}. We may apply a separation of variables to equation \eqref{Berry 2} to get a set of energy eigenmodes $\phi_{n}(\rho,t)=u_{n}(\rho,t)\exp\left(-i\mathcal{E}_{n}\hbar\right)$ such that
\begin{equation}
\label{Berry 4}
\mathcal{E}_{n}u_{n}=
\frac{-\hbar^{2}}{2m}\frac{\partial ^{2}u_{n}}{\partial \rho^{2}} + \left(V(\rho)+\frac{1}{2}k\rho^{2}\right)u_{n}
\end{equation}
Equation \eqref{Berry 4} then results in a set of expanding eigenmodes in the original coordinates:
\begin{equation}
\psi_{n}(x,t)=\frac{u_{n}(x/l(t))}{l(t)}\exp\left(\frac{i}{\hbar}\left(\frac{ml'x^{2}}{2l}-\mathcal{E}_{n}\tau(t)\right)\right)
\end{equation}
Note that the eigenmode $u_{n}(\rho,\tau)$ experiences a potential $V(\rho)$ with an additional harmonic term $k\rho^{2}/2$. If $k=0$, then by equation \eqref{Berry 3.5} $b=\sqrt{ac}$ and
\begin{equation}
l(t)=\sqrt{at^{2}+2bt+c}=\sqrt{at^{2}+2\sqrt{ac}t+c}=\sqrt{a}t+\sqrt{c}
\end{equation}
which corresponds to Niederer's expansion in appendix \ref{Niederer appendix}

This method is particularly useful when $V(\rho)$ only takes on the values $0$ or $\infty$, in which case $\alpha(l)$ becomes redundant and equation \eqref{Berry 1} is just the free space Schrodinger equation with some boundary conditions. Thus Berry and Klein's method fully generalises Greenberger's method to arbitrary potential wells. Berry and Klein comment that for an expanding box and $k=0$, ``$u_{n}$ is a simple trigonometric function". In this case, the origin of expansion ($b$ in figure \ref{fig:1}) can be chosen so as  to control $v_{1}$ and $v_{2}$.

\section{Revivals in an infinite potential well with walls moving at constant velocities}
\label{Linear revivals}

In this section we will examine the phenomenon of revivals in the infinite potential well, and extend this treatment to a well with non-accelerating walls.

The solutions $\psi_{n}(x,t)$ from equation \eqref{The solution set} are orthonormal for each $t$:
\begin{equation}
    \int_{w_{1}(t)}^{w_{2}(t)}\psi_{m}^{*}(x,t)\psi_{n}(x,t)\,dx=\delta_{nm}
\end{equation}
This follows immediately from the time independent case as the phase term $\exp\left(i\theta(x,t)\right)$ cancels by conjugation. The solutions are complete, which follows from Fourier theory. To expand $\psi(x,0)$ as a sum of $\psi_{n}(x,0)$, the summation coefficients are given by
\begin{equation}
\begin{aligned}
a_{n}&=\int_{0}^{w_{0}}\psi_{n}^{*}(x,0)\psi(x,0)\,dx \\
&=\sqrt{\frac{2}{w_{0}}}\int_{0}^{w_{0}}\left(\sin\left(\frac{n\pi x}{w_{0}}\right)\exp(-i\theta(x,t))\right)\psi(x,0)\,dx \\
&=\sqrt{\frac{2}{w_{0}}}\int_{0}^{w_{0}}\sin\left(\frac{n\pi x}{w_{0}}\right)\left(\exp(-i\theta(x,t))\psi(x,0)\right)\,dx
\end{aligned}
\end{equation}
and thus the time evolution of $\psi(x,t)$ may be determined:
\begin{equation}
\label{Revivals 1}
\psi(x,t)=\sum_{n=1}^{\infty}a_{n}\psi_{n}(x,t)
\end{equation}

Rather than analysing the sum \eqref{Revivals 1}, we will apply the Greenberger transformation \eqref{Greenberger transformation} to $\psi(x,t)$ and analyse $\phi(y,\tau)$ instead. The time evolution of a particle in a box is well known \cite{Berry:2001}, but we will briefly recount it here. The eigenmode expansion of $\phi(y,\tau)$ is
\begin{equation}
\label{Phi eigenmode expansion}
\begin{aligned}
\phi(y,\tau)&=\sum_{n=1}^{\infty}c_{n}\sqrt{2}\sin(n\pi y)\exp\left(-i\frac{n^{2}\pi^{2}\hbar \tau}{2m}\right) \\
&=\sum_{n=1}^{\infty}c_{n}\sqrt{2}\sin(n\pi y)\exp(-in^{2}\pi\tau ')
\end{aligned}
\end{equation}
where 
\begin{equation}
\label{linear tau'}
\tau'=\frac{\hbar\pi}{2m}\tau=\frac{\hbar\pi}{2m}\int_{0}^{t}\frac{dt'}{w^{2}(t)}=\frac{\hbar\pi}{2m\Delta v}\left(\frac{1}{w(0)}-\frac{1}{w(t)}\right)
\end{equation}
and
\begin{equation}
\label{Phi eigenmode coefficients}
c_{n}=\int_{0}^{1}\sqrt{2}\sin(n\pi y)\phi(y,0)\,dy
\end{equation}
In order to relate the initial wavefunction $\phi(y,0)$ to the wavefunction $\phi(y,\tau ')$ at an arbitrary time later, consider the propagator $U(y,z,\tau ')$:
\begin{equation}
    \phi(y,\tau ')=\int_{0}^{1}U(y,z,\tau ')\phi(z,0)\,dz
\end{equation}
The propagator may be calculated directly by substituting equation \eqref{Phi eigenmode coefficients} into equation \eqref{Phi eigenmode expansion}:
\begin{equation}
\begin{multlined}
U(y,z,\tau ')=2\sum_{n=1}^{\infty}\sin\left(n\pi y\right)\sin\left(n\pi z\right)\exp\left(-i\pi n^{2}\tau '\right) \\
\begin{aligned}
&=\frac{-1}{2}\sum_{n=1}^{\infty}\left(e^{in\pi y}-e^{-in\pi y}\right)\left(e^{in\pi z}-e^{-in\pi z}\right)e^{-i\pi n^{2}\tau '} \\
&=\frac{-1}{2}
\sum_{n=1}^{\infty}\left(e^{in\pi(y+z)}+e^{-in\pi(y+z)}-e^{in\pi(y-z)}-e^{-in\pi(y-z)}\right)e^{-i\pi n^{2}\tau '} \\
&=\frac{1}{2}
\sum_{n=-\infty}^{\infty}\left(e^{in\pi(y-z)}-e^{in\pi(y+z)}\right)e^{-i\pi n^{2}\tau '}\\
&=\frac{1}{2}\left(\vartheta\left(\frac{y-z}{2},-\tau '\right)-\vartheta\left(\frac{y+z}{2},-\tau '\right)\right)
\end{aligned}
\end{multlined}
\end{equation}
where \emph{Jacobi's elliptic theta function} $\vartheta(y,\tau ')$ \cite{Apostol} is defined as
\begin{equation}
\vartheta(y,\tau ')=\sum_{n=-\infty}^{\infty}e^{2\pi in y+\pi i n^{2}\tau '}
\end{equation}
Thus
\begin{equation}
\label{theta propagation}
\phi(y,\tau ')=\frac{1}{2}\int_{0}^{1}\left(\vartheta\left(\frac{y-z}{2},-\tau '\right)-\vartheta\left(\frac{y+z}{2},-\tau '\right)\right)\phi(z,\tau ')\,dz 
\end{equation}

Equation \eqref{theta propagation} takes on a form very similar to a convolution, and so the dynamics of a wavefunction for a particle in a box depend quite explicitly on Jacobi's elliptic function. $\vartheta(y,\tau ')$ is a very interesting function, with many subtle properties which we will not be able to fully review here. Most importantly for our present purposes,
$\vartheta(y,\tau ')$ may be calculated explicitly for rational values of $\tau '$:
\begin{equation}
\label{Rational theta}
\vartheta\left(y,\frac{p}{q}\right)=\sum_{s=0}^{2q-1}c_{s}(p,2q)\delta\left(y-\frac{s}{2q}\right)
\end{equation}
where $c_{s}(p,2q)$ is a \emph{generalised quadratic Gauss sum}.
\begin{equation}
c_{s}(p,2q)=\frac{1}{2q}\sum_{r=0}^{2q-1}e^{2\pi i\left((p/2q)r^{2}+(s/2q)r\right)}
\end{equation}
Generalised Gauss sums satisfy some very nice arithmetic relations which allow for their calculation \cite{Apostol}. 
In particular, when $s$ and $q$ have the same parity:
\begin{equation}
\label{c(1,2q)}
c_{s}(1,2q)=\frac{1}{\sqrt{q}}e^{\frac{i\pi}{4}(1-s^{2}/q)}
\end{equation}
and $c_{s}(1,2q)=0$ otherwise. Interestingly, equation \eqref{c(1,2q)} follows from the Appell transformation \eqref{Appell transformation}. In particular, this gives us an explicit formula for $\vartheta(x',1/q)$:
\begin{equation}
\label{theta 1/q}
\vartheta\left(y,\frac{1}{q}\right)=\frac{1}{\sqrt{q}}\sum_{\substack{s=0 \\ s \equiv q\left(\textrm{mod}\,2\right)}}^{2q-1}e^{\frac{i\pi}{4}\left(1-s^{2}/q\right)}\delta\left(y-\frac{s}{2q}\right)
\end{equation}
Further, observe that $\vartheta(y,-\tau ')=\vartheta(y,\tau ')^{\ast}$. Thus equation \eqref{theta propagation} combined with \eqref{Rational theta} fully describes $\phi(y,\tau ')$ for rational values of $\tau '$; $\phi(y,\tau ' = p/q)$ is a sum of translates of the initial field $\phi(y,0)$:
\begin{equation}
\label{Phi propagation}
\begin{multlined}
\phi\left(y,\tau'=\frac{p}{q}\right)\\
\begin{aligned}
&=\sum_{s=0}^{2q-1}\frac{c_{s}(p,2q)^{\ast}}{2}\int_{0}^{1}\left(\delta\left(\frac{y-z}{2}-\frac{s}{2q}\right)-\delta\left(\frac{y+z}{2}-\frac{s}{2q}\right)\right)\phi(z,0)\,dz \\
&=\sum_{s=0}^{2q-1}c_{s}(p,2q)^{\ast}\left(\phi\left(y-\frac{s}{q},0\right)-\phi\left(-y+\frac{s}{q},0\right)\right)
\end{aligned}
\end{multlined}
\end{equation}
Note that it's possible $y-s/q\notin[0,1]$ even when $y\in[0,1]$, which implies $\phi(y-s/q,\tau)$ is ill defined. The trick is to extend $\phi(y,\tau)$ to be defined for all $y$ by using equation \eqref{Phi eigenmode expansion}. This is equivalent to extending $\phi(y,\tau)$ to a periodic function with period $2$ such that $\phi(-y,\tau)=-\phi(y,\tau)$. Each of the translates of $\phi$ in equation \eqref{Phi propagation} is known as a \emph{revival} of the original wavefunction \cite{AronsteinStroud}. 

\begin{figure}[h]
    \def\svgwidth{\columnwidth}
    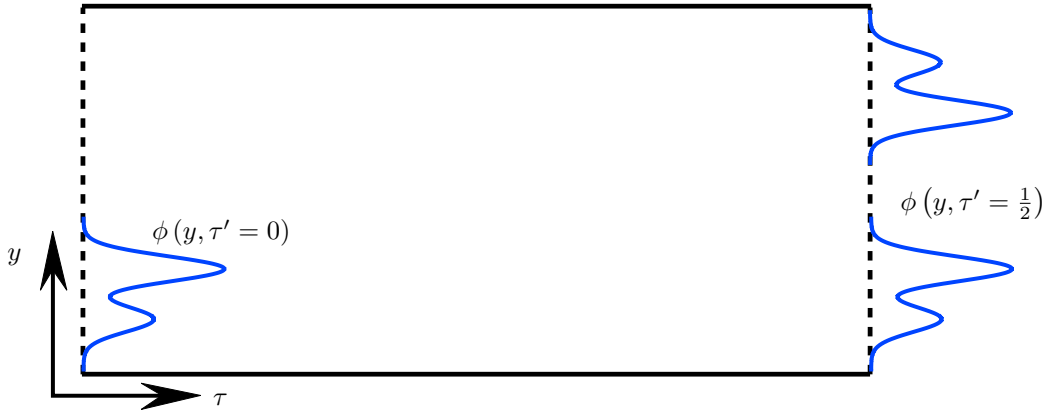
    \caption{The two revivals of a wavefunction in an infinite potential well with fixed walls at $\tau'=1/2$. Note that one of the revivals is a reflection of the original wavefunction.}
    \label{fig:double revival}
    \centering
\end{figure}

As an example of equations \eqref{Phi propagation} and \eqref{c(1,2q)}, we may evaluate $\phi(y,\tau'=1/2)$:
\begin{equation}
\label{Phi double revivals}
\begin{aligned}
\phi\left(y,\tau'=\frac{1}{2}\right) &= \frac{1}{\sqrt{2}}\sum_{s=0,2}e^{\frac{i\pi}{4}\left(s^{2}/q-1\right)}\left(\phi\left(y-\frac{s}{2},0\right)-\phi\left(-y+\frac{s}{2},0\right)\right) \\
& \approx\frac{1}{\sqrt{2}}\left(e^{\frac{-i\pi}{4}}\phi(y,0)-e^{\frac{i\pi}{4}}\phi(1-y,0)\right) \\
& =\frac{1}{\sqrt{2}}\left(e^{\frac{-i\pi}{4}}\phi(y,0)+e^{\frac{i5\pi}{4}}\phi(1-y,0)\right)
\end{aligned}
\end{equation}
We have dismissed $\phi(y-1)$ and $\phi(-y)$ as being negligible for $y\in[0,1]$. This is true provided $\phi(y,0)$ is narrow enough and far enough away from the well boundaries. Equation \eqref{Phi double revivals} is sketched schematically in figure \ref{fig:double revival}. Note that the two revivals at $\tau'=1/2$ are symmetrically located about $y=1/2$, and that they are mirror images of each other. 

In general, we expect that at $\tau'=p/q$ (where $p$ and $q\neq 0$ are integers with no common divisors) there will be $q$ revivals of $\phi(y,0)$. Sometimes the revivals will overlap perfectly, in which case there will appear to be less than $q$ revivals. A similar phenomenon of revivals occurs in optics, known as the Talbot effect, which is particularly useful in photonics. The revival of source images via the Talbot effect and their combinations is the key feature in a class of photonic devices known as multimode interferometers \cite{Cooney}.

\begin{figure}[t]
    \includegraphics[width=\textwidth]{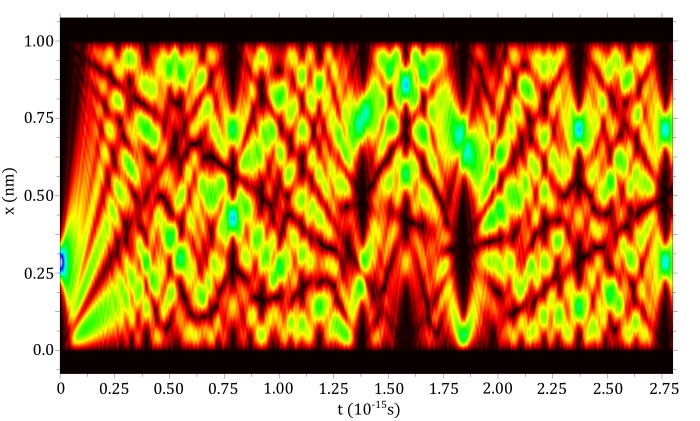}
    \caption{An electron evolving in a $1$nm box.}
    \centering
    \label{fig:straight simulation}
\end{figure}

Figure \ref{fig:straight simulation} provides a numerical simulation of $\left|\phi(x,t)\right|^{2}$, where the double revival occurs at about $2.75 \times 10^{-15} s$. This revival time may be calculated by solving for $t(\tau')$:
\begin{equation}
\begin{gathered}
\tau'=\frac{h\pi\tau}{2m}=\frac{h\pi t}{2mw^{2}} \\
\Rightarrow t = \frac{2m w^{2}}{\hbar \pi}\tau'
\end{gathered}
\end{equation}
Setting $\tau'=1/2$ and $m$ to the electron mass yields the time for a double revival. Interestingly,  $w=10$ cm $\Rightarrow t/t'\approx 30$ s, which would make a very interesting toy system if an electron could be seen by the naked eye.

Notice that other distinct revivals occur in figure \ref{fig:straight simulation}. This is to be expected by \eqref{Phi propagation}. However, revivals don't occur for every rational $\tau'$. This is because the initial wavefunction $\phi(x,0)$ has a finite width. If the width of the initial wavefunction is $1/10$ the width of the well, then clearly we will not be able to see $10$ revivals clearly. Thus, for more primitive fractions like $1/4,1/3$ etc. the revivals are much more clear. As the width of the initial wavefunction decreases, more revivals become visible. For small enough widths, the wavefunction will eventually tend towards a fractal shape known as a \emph{Talbot carpet} \cite{Berry:2001}. Conveniently, the summation of $q$ overlapping revivals at $\tau'=1/q$ for large $q$ closely resembles free space propagation.

By applying an inverse Greenberger transformation to $\phi(y,\tau)$ and $\phi(y,0)$ in equation \eqref{Phi propagation}, we may derive an expression for $\psi(x,t)$ in terms of $\psi(x,0)$: 
\begin{multline}
\label{Psi propagation x}
\psi\left(x,t\left(\tau'=\frac{p}{q}\right)\right)=\sqrt{\frac{w_{0}}{w(t)}}\exp\left(i\theta(x,t)\right)
\sum_{s=0}^{2q-1}c_{s}(p,2q)^{\ast} \\
\left(\exp\left(-i\theta\left(\frac{w(0)}{w(t)}(x-v_{1}t)-\frac{s}{q}w(0),0\right)\right)\psi\left(\frac{w(0)}{w(t)}(x-v_{1}t)-\frac{s}{q}w(0),0\right)-\right. \\
\left.\exp\left(-i\theta\left(\frac{w(0)}{w(t)}(-x+v_{1}t)+\frac{s}{q}w(0),0\right)\right)\psi\left(\frac{w(0)}{w(t)}(-x+v_{1}t)+\frac{s}{q}w(0),0\right)\right)
\end{multline}
Equation \eqref{Psi propagation x} is much simpler if we use the normalised coordinate $y$:
\begin{multline}
\label{Psi propagation y}
\psi\left(y,t\left(\tau'=\frac{p}{q}\right)\right)=\sqrt{\frac{w_{0}}{w(t)}}e^{i\theta(y,t)}
\sum_{s=0}^{2q-1}c_{s}(p,2q)^{\ast} \\
\left(e^{-i\theta\left(y-s/q,0\right)}\psi\left(y-\frac{s}{q},0\right)-
e^{-i\theta\left(-y+s/q,0\right)}\psi\left(-y+\frac{s}{q},0\right)\right)
\end{multline}
$\psi(y-s/q,t)$ may be ill defined, as above. However in this case we must extend the product $e^{-i\theta}\psi(y,0)=\phi/\sqrt{w}$ by periodicity and antisymmetry, not just $\psi(y,0)$. 

\begin{figure}[h]
    \def\svgwidth{\columnwidth}
    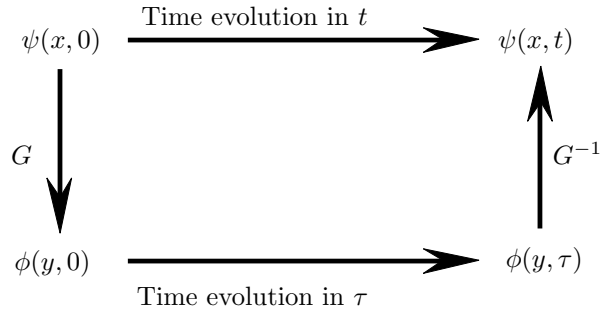
    \caption{The Greenberger transformation makes $t$ evolution of $\psi(x,t)$ and $\tau$ evolution of $\phi(y,\tau)$ equivalent.}
    \label{fig:Greenberger commutative diagram}
    \centering
\end{figure}
 
In practice, to deduce the time evolution of $\psi(x,t)$ it is easier to move to $\phi(y,\tau)$ and deduce the time evolution there. This is expressed graphically in figure \ref{fig:Greenberger commutative diagram}.

\begin{figure}[h]
    \def\svgwidth{\columnwidth}
    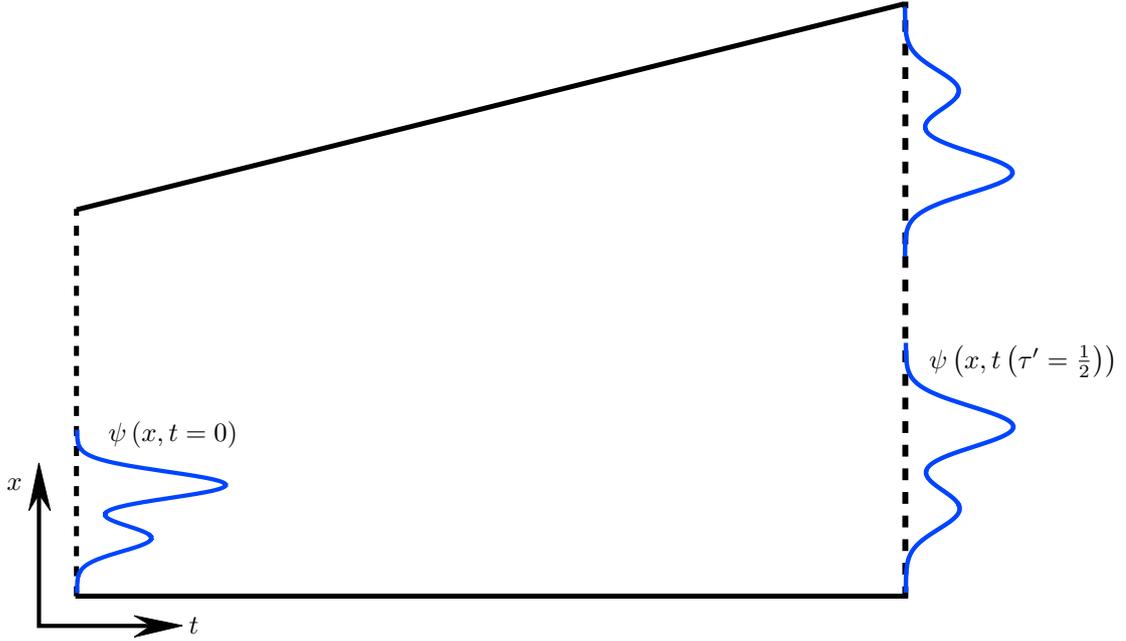
    \caption{An expanding well with one wall moving at constant velocity, and a wavefunction experiencing a double revival}
    \label{fig:expanding double revival}
    \centering
\end{figure}

Equations \eqref{Psi propagation x} and \eqref{Psi propagation y} show that the revivals of $\psi(x,t)$ differ from those of $\phi(y,\tau)$ in 3 main ways. Firstly, he width of the revivals at time $t$ is $w(t)/w(0)$ times the width of the original wavefunction. Secondly, the amplitude of the revivals is $\sqrt{w(0)/w(t)}$ times the amplitude of the original wavefunction. Thirdly, the phase of each of the revivals will be altered by a phase term coming from the Greenberger transformation. However, the location of the revivals relative to the width of the well are the same. A schematic of a linearly expanding well with a double revival is shown in figure \ref{fig:expanding double revival}. 

\begin{figure}[t]
    \includegraphics[width=\textwidth]{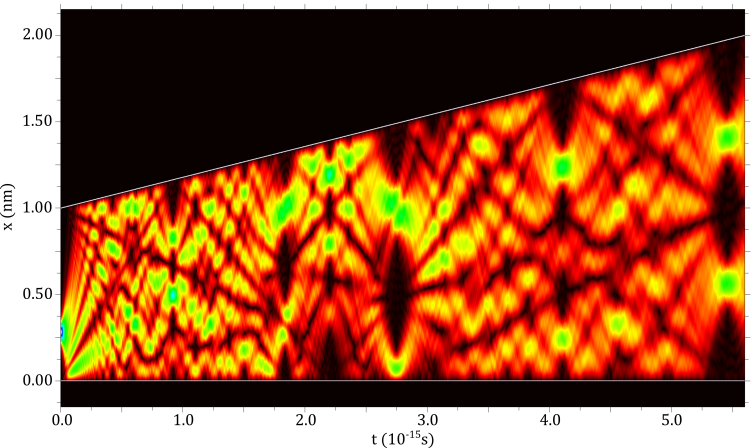}
    \caption{A numerical simulation of an electron in a linearly expanding well with one moving wall.}
    \centering
    \label{fig:linear simulation}
\end{figure}

Figure \ref{fig:linear simulation} displays a numerical simulation for $|\psi(x,t)|^{2}$ in a well with one wall moving at a constant velocity. The initial width of the well is $w(0)=1$ nm, and the velocity of the moving wall $v$ has been chosen that when a double revival occurs the width of the well is $w(t)=2w(0)=2$ nm. By equation \eqref{linear tau'},
\begin{gather}
    \tau'= \frac{1}{2} =\frac{\hbar\pi}{2m v}\left(\frac{1}{2w(0)}\right) \\
    \Rightarrow v = \frac{\hbar \pi}{2mw(0)} = 1.82 \times 10^{5}\,\textrm{ms}^{-1} \\
\Rightarrow t = \frac{w(t)-w(0)}{v} =\frac{2mw(0)^{2}}{\hbar\pi} = 5.5\times10^{-15} s 
\end{gather}

The input location is the same in both figures \ref{fig:straight simulation} and \ref{fig:linear simulation}. The result is that the two time evolution patterns are identical up to a stretching, as follows from equation \eqref{Psi propagation y}. As the width of the well widens, the revivals become wider and the interval between revivals becomes larger. Both of these features are clearly visible in figure \ref{fig:linear simulation}.

Expanding on the previous calculation, we may solve for $\tau(t')$ explicitly. By equations \eqref{tau(t)} and \eqref{t(tau)}
\begin{gather}
\label{tau'(t)}
\tau'(t)=\frac{\hbar \pi}{2m}\frac{t}{w_{0}^{2}+\Delta v w_{0}t} \\
\label{t(tau')}
t(\tau')=\frac{w_{0}^{2}\tau'}{(h\pi/2m)-w_{0}\Delta v\tau'}
\end{gather}

\begin{figure}[h]
    \def\svgwidth{\columnwidth}
    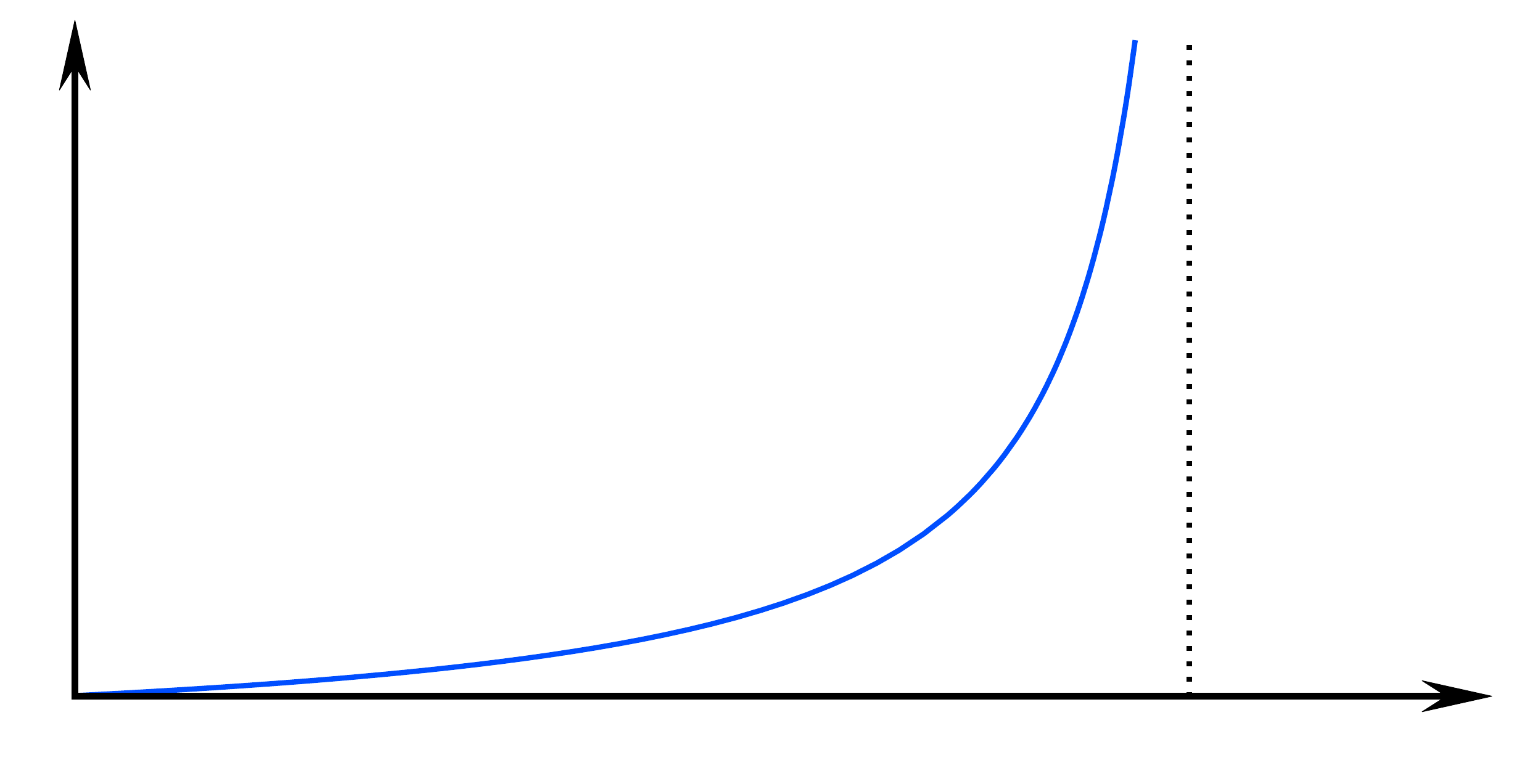
    \caption{A graph of $t(\tau')$ for an potential well expanding with constant velocity $\Delta v$.}
    \label{fig:t(tau') linear}
    \centering
\end{figure}

The behaviour of equations \eqref{tau'(t)} and \eqref{tau'(t)} splits into two cases, the contracting $(\Delta v < 0)$ and expanding $(\Delta v > 0)$ well. In the following we will consider positive times $t>0$, but changing the sign of $t$ just exchanges the roles of contraction and expansion. In both cases, $t(\tau=0)=0$.

If $\Delta v >0$, then $t(\tau')$ increases monotonically until it diverges at $\tau = h\pi/(2mw_{0}\Delta v)$. Equivalently,
\begin{equation}
\lim_{t\to\infty} \tau '(t)=\frac{\hbar\pi}{2mw_{0}\Delta v}
\end{equation}
Thus in contrast to the case with fixed walls, there is an upper bound on the value of $\tau '$ that can be achieved in a finite time when the well is expanding. This means that certain revivals will not be possible. As $t$ tends to infinity, the rate of increase of $\tau'(t)$ drops off rapidly. The result is that for large $t$ $\psi(x,t)$ will appear to almost stop evolving, however the form of $\psi(x,t)$ will still stretch in proportion to the width of the well.

If $\Delta v < 0$, $t$ is finite for positive $\tau '$ but $\tau'$ diverges for $t=-w_{0}/\Delta v$. Equivalently
\begin{equation}
\lim_{\tau '\to\infty}t(\tau ')=-\frac{w_{0}}{\Delta v}
\end{equation} Of course, this is the time when the walls of the well collide. Thus for a contracting well, all possible values of $\tau'$ are attainable and all such revivals will occur within a finite time $-w_{0}/\Delta v$.

Notice that the deductions of the above two paragraphs are invariant (up to a scaling) under the action of swapping $t$ and $\tau '$ whilst also swapping the roles of contraction and expansion. In other words, the map $(t,\tau ')\mapsto (-\tau ', -t)$ is a sort of symmetry of the problem. This follows from the group structure of expansions \eqref{expansion transformation}. In the language of the Niederer transformations from appendix \ref{Niederer appendix}, $\left[\alpha - \alpha\right]=[0]$ is the identity map, so $\left[-\alpha\right]=\left[\alpha\right]^{-1}$ or equivalently $\left[-\alpha\right]^{-1}$ is the identity map. But up to a scaling, $\left[-\alpha\right]^{-1}$ is just $(t,\tau ')\mapsto (-\tau ', -t)$.

\section{The infinite potential well with slowly accelerating walls}
\label{Nonlinear box and revivals}

In this section we will extend the analysis of the previous two sections to infinite potential wells with walls that do not necessarily move at a constant velocity. Thus, consider a wavefunction $\psi(x,t)$ obeying Schrodinger's equation \eqref{Schrodinger's equation} subject to the boundary conditions $\psi(w_{1}(t),t)=\psi(w_{2}(t),t)=0$, where $w_{1}(t)$ and $w_{2}(t)$ are the positions of the lower and upper walls respectively. As before we will attempt to solve this problem using transformation methods in free space; the two transformations we require are the extended Galilean and Greenberger transformations.

\subsection{Extended Galilean Transformations}

\begin{figure}[h]
    \def\svgwidth{\columnwidth}
    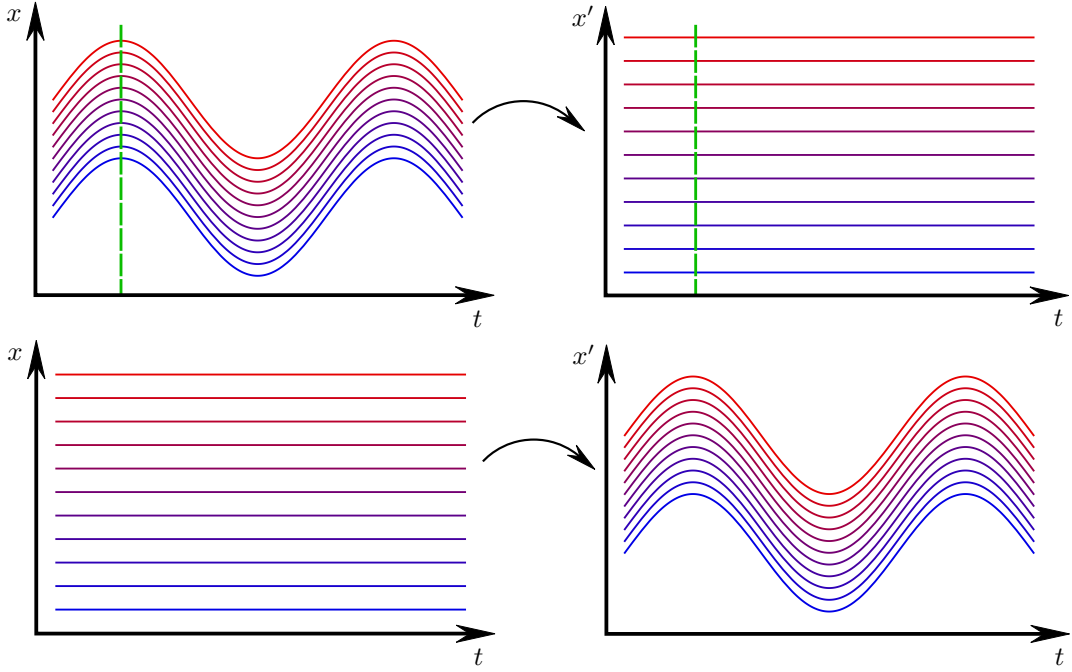
    \caption{Two possible interpretations of an extended Galilean transformation for sinusoidal $d(t)$.}
    \label{fig:EGT coordinates}
    \centering
\end{figure}

An \emph{extended Galilean transformation} \cite{Rosen} is a transformation of the form $(x,t)\mapsto(x-d(t),t)$ where $d(t)$ is a time dependent displacement. Thus let $x'=x-d(t)$. Figure \ref{fig:EGT coordinates} illustrates two possible interpretations of this transformation. In the top half of the figure, the set of curves $x-d(t)$ for varying $x$ are mapped to horizontal lines in the $(x',t)$ plane. Alternatively, the set of horizontal lines in the $(x,t)$ plane are mapped to the curves $x'+d(t)$ in the $(x',t)$. We will mainly be interested in the first interpretation, as we wish to simplify non-trivial geometries. These two interpretations are inverse to each other. 

If $\psi(x,t)$ is a solution to equation \eqref{Schrodinger's equation}, then by the same procedure as in section \ref{GB section} $\psi(x',t)$ is a solution to
\begin{equation}
\label{EGT 1}
i\hbar\left(\frac{\partial \psi}{\partial t}-\dot{d(t)}\frac{\partial \psi}{\partial x'}\right)=-\frac{\hbar^{2}}{2m}\frac{\partial ^{2}\psi}{\partial x'^{2}}
\end{equation}
Let $\phi=\psi\exp\left(-i\theta(x',t)\right)$, for some function $\theta$. Equation \eqref{EGT 1} becomes
\begin{equation}
\label{EGT 2}
i\hbar\frac{\partial \phi}{\partial t} = 
\frac{-\hbar^{2}}{2m}\frac{\partial ^{2} \phi}{\partial x'^{2}} 
- i\hbar\left(\frac{\hbar \theta_{x'}}{m}-\dot{d}\right)\frac{\partial\phi}{\partial x'}
+\hbar\left(\theta_{t}-\dot{d}\theta_{x'}+\frac{\hbar}{2m}\left(\theta_{x'x'}+\theta_{x'}^{2}\right)\right)\phi
\end{equation}
To ensure the dissipation term $\partial \phi /\partial x'$ vanishes, we set 
\begin{gather}
\frac{\hbar\theta_{x'}}{m}=\dot{d} \\
\Rightarrow \theta(x',t)=\frac{m\dot{d}x'}{\hbar}+c(t)
\end{gather}
where $c(t)$ is some function of $t$ (i.e. independent of $x'$). The coefficient of $\phi$ in equation \eqref{EGT 2} is then
\begin{equation}
\label{EGT 4}
m\ddot{d}x'+\hbar\dot{c(t)}-\frac{m\dot{d}^{2}}{2}
\end{equation}
To simplify equation \eqref{EGT 4}, we may set $\dot{c}=m\dot{d}^{2}/2\hbar$. Therefore
\begin{gather}
c(t)=\frac{m}{2\hbar}\int_{0}^{t}\dot{d(t')}^{2}\,dt' \\
\label{EGT 5}
\Rightarrow \theta(x',t)=\frac{m}{\hbar}\left(\dot{d(t)}x'+\frac{1}{2}\int_{0}^{t}\dot{d(t')}^{2}\,dt'\right)
\end{gather}
and $\phi(x',t)$ satisfies
\begin{equation}
i\hbar\frac{\partial \phi}{\partial t}=-\frac{\hbar^{2}}{2m}\frac{\partial ^{2} \phi}{\partial x'^{2}}+\left(m\ddot{d(t)}x'\right)\phi
\end{equation}
More generally, if $\psi(x,t)$ is subjected to a potential $V(x,t)$ then under the same transformation $\phi(x',t)$ satisfies
\begin{equation}
\label{EGT Schrodinger equation}
    i\hbar\frac{\partial \phi}{\partial t}=-\frac{\hbar^{2}}{2m}\frac{\partial ^{2} \phi}{\partial x'^{2}}+\left(V\left(x'-d(t)\right)+m\ddot{d(t)}x'\right)\phi
\end{equation}
Equation \eqref{EGT Schrodinger equation} makes it clear that the extended Galilean transformation induces a change on the potential 
\begin{equation}
\label{EGT V transformation}
V(x,t)\mapsto V'(x',t)=V\left(x'-d(t)\right)+m\ddot{d(t)}x'
\end{equation}
The additional $m\ddot{d}x'$ term is the potential corresponding to a uniform force $m\ddot{d}$. This is the same classical fictitious force that occurs in any non-inertial reference frame. Note that $V'$ may be a time dependent potential even if $V$ is not.

\begin{figure}[h]
    \def\svgwidth{\columnwidth}
    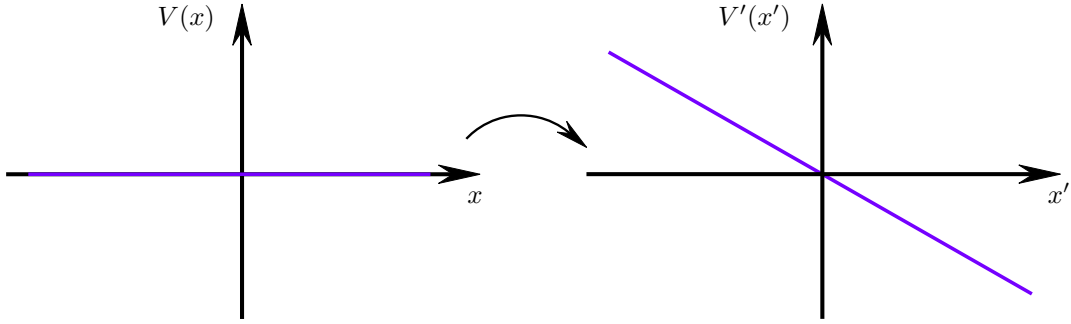
    \caption{The induced potential of an extended Galilean transformation.}
    \label{fig:Induced linear potential}
    \centering
\end{figure}

Figure \ref{fig:Induced linear potential} shows this change for a constant potential $V(x)$ along the green dashed line in figure \ref{fig:EGT coordinates}. In the $(x',t)$ plane, $V'$ pushes the wavefunction towards higher values (in red) of $x'$. This can be understood from the $(x,t)$ picture also. A wavefunction evolving from the green line will evolve homogeneously, without any $x$ dependence. However due to the geometry of the sinusoids, the red curves sweep down in front of this evolving wavefunction. The result is that $(x',t)$ plane, the wavefunction moves towards the red region, the cause of which may be experienced as a force.

Rosen derived the extended Galilean transformations using an argument based on Einstein's principle of equivalence \cite{Rosen}. Holstein derived equation \eqref{EGT 5} using a path integral argument \cite{Holstein}. More recently, Klink has derived the extended Galilean transformations by interpreting $\theta(x,t)$ in equation \eqref{EGT 5} as a generating function \cite{Klink,MacGregor}. In appendix \ref{Galilean group appendix}, the group structure of extended Galilean transformations is discussed.

\subsection{Extended Greenberger Transformations}

\begin{figure}[h]
    \def\svgwidth{\columnwidth}
    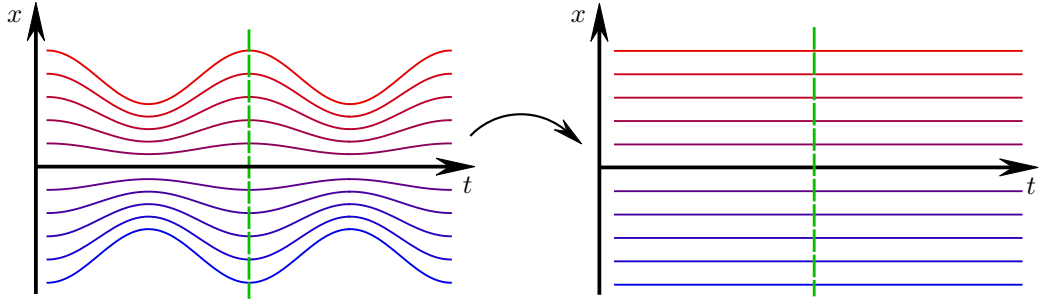
    \caption{An extended Greenberger transformation of the form $y=x/w(t)$.}
    \label{fig:Nonlinear Greenberger}
    \centering
\end{figure}

An \emph{extended Greenberger transformation} is a Greenberger transformation as in section \ref{GB section} with a scaling term $w(t)>0$ such that $\ddot{w(t)}\neq 0$. To begin with, we will consider transformations that preserve the origin, i.e. transformations of the form $y=x/w(t)$. One such transformation is shown in figure \ref{fig:Nonlinear Greenberger}. Define $\phi(y,t)$ as in equation \eqref{Greenberger 2.5}; 
\begin{equation}
\phi(y,t)=\sqrt{w(t)}\psi(y,t)\exp\left(-i\theta(y,t)\right)
\end{equation}
with
\begin{equation}
\label{EGBT theta 0}
\theta(y,t)=\frac{m\dot{w(t)}w(t)y^{2}}{2\hbar}
\end{equation}
Greenberger \cite{Greenberger} showed that if $\psi(x,t)$ is subject to a potential $V(x/w(t))$, then $\phi(y,\tau)$ satisfies the following Schrodinger equation
\begin{equation}
\label{EGBT Schrodinger equation potential}
i\hbar\frac{\partial\phi}{\partial\tau}=-\frac{\hbar^{2}}{2m}\frac{\partial^{2}\phi}{\partial y^{2}}+\left(w(t)^{2}V(y)+\frac{m}{2}w(t)^{3}\ddot{w(t)}y^{2}\right)\phi
\end{equation}
where
\begin{equation}
\label{EGBT tau}
\tau(t)=\int_{0}^{t}\frac{1}{w(t')^{2}}\,dt'
\end{equation}
If $V\left(x/w(t)\right)=0$, then equation \eqref{EGBT Schrodinger equation potential} reduces to
\begin{equation}
\label{EGBT Schrodinger equation}
i\hbar\frac{\partial\phi}{\partial\tau}=-\frac{\hbar^{2}}{2m}\frac{\partial^{2}\phi}{\partial y^{2}}+\left(\frac{m}{2}w(t)^{3}\ddot{w(t)}y^{2}\right)\phi
\end{equation}
While an extended Galilean transformation induces a linear potential, an extended Greenberger transformation induces a quadratic potential $mw^{3}\ddot{w}y^{2}/2$. Inverting this argument, we may think of a wavefunction in a harmonic oscillator as the result of a wavefunction evolving in ``stretched" free space. The spring constant $mw^{3}\ddot{w}y^{2}/2$ is time independent when $w(t)=\sqrt{at^{2}+2bt+c}$, which is the case studied by Berry and Klein \cite{Berry:1984}.

\begin{figure}[h]
    \def\svgwidth{\columnwidth}
    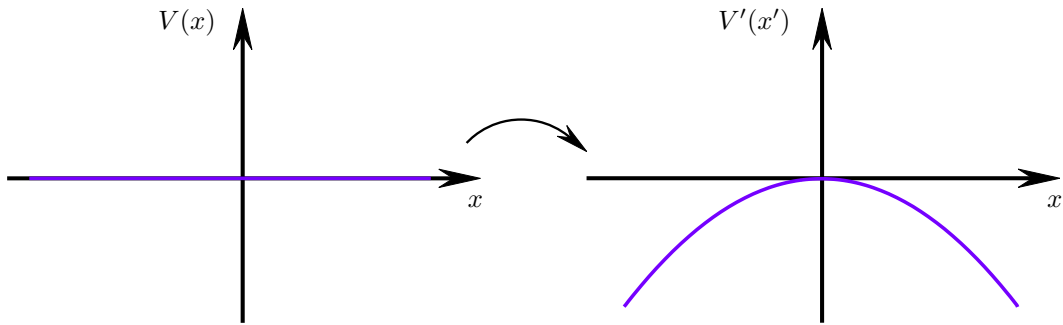
    \caption{A quadratic potential $V'(x')$ induced by an extended Greenberger transformation.}
    \label{fig:Induced quadratic potential}
    \centering
\end{figure}

Figure \ref{fig:Induced quadratic potential} shows the effect of the extended Greenberger transformation on a constant potential $V$ at the green dashed line. The induced potential $V'(x')$ is a negative quadratic, pushing the wavefunction away from the origin. As in the Galilean case, this pushing may be interpreted geometrically in the $(x,t)$ plane. 

The extended Greenberger transformations first appeared in an appendix of Hill and Wheeler discussing nuclear fission \cite{HillWheeler}. Greenberger then applied this transformation to particles in a box with one wall moving at a constant velocity \cite{Greenberger}. Takagi published a series of papers  discussing in detail both the extended Galilean and Greenberger transformations \cite{Takagi:1,Takagi:2,Takagi:3}. In these papers Takagi investigates how said transformations alter the electromagnetic potentials $\vec{A}$ and $V$, and how to use these transformations to solve problems involving electromagnetism and potentials of a harmonic or linear nature. Takagi refers to the variable $y$ as a ``comoving frame". 

Extended Greenberger transformations also form a group; scaling $x$ by $1/w_{1}(t)$ and then $1/w_{2}(t)$ is the same as scaling by $1/w_{1}(t)w_{2}(t)$. However this group structure is harder to analyse than the Galilean case as the transformations affect both the spatial and temporal variables.

The overdots in equation \eqref{EGBT Schrodinger equation} denote differentiation with respect to $t$. We may re-express $\ddot{w(t)}$ in terms of $\tau$ using the chain rule
\begin{equation}
\ddot{w(t)}
= \frac{d}{dt}\left(\frac{dw}{dt}\right)
=\frac{1}{w^{2}}\frac{d}{d\tau}\left(\frac{1}{w^{2}}\frac{dw}{d\tau}\right)
=\frac{1}{w^{4}}\frac{d^{2}w}{d\tau^{2}}-\frac{2}{w^{5}}\left(\frac{dw}{d\tau}\right)^{2}
\end{equation}
Thus, equation \eqref{EGBT Schrodinger equation} is equivalent to
\begin{equation}
\label{EGBT Schrodinger equation tau}
i\hbar\frac{\partial\phi}{\partial\tau}
=-\frac{\hbar^{2}}{2m}\frac{\partial^{2}\phi}{\partial y^{2}}
+\frac{m}{2}\left(\frac{\ddot{w(\tau)}}{w(\tau)}-2\left(\frac{\dot{w(\tau)}}{w(\tau)}\right)^{2}\right)y^{2}\phi
\end{equation}
where overdots in equation \eqref{EGBT Schrodinger equation tau} denote differentiation with respect to $\tau$. In future, we will interpret an overdot as denoting differentiation with respect to the variable being used as the function argument. If no argument is present, we will use $t$ by default. 

Let us now consider the more general Greenberger transformation $y=(x-d(t))/w(t)$ applied to a wavefunction $\psi(x,t)$ in free space. This transformation is a result of first translating by $d(t)$ and then scaling by $1/w(t)$. If $x'=x-d(t)$ and $\phi'(x',t)=\psi(x',t)\exp(-i\theta_{T}(x',t))$ with $\theta_{T}(x',t)$ defined as in equation \eqref{EGT 5}, then
\begin{equation}
i\hbar\frac{\partial \phi '}{\partial t}=-\frac{\hbar^{2}}{2m}\frac{\partial^{2} \phi '}{\partial x'^{2}}+m\ddot{d}x'
\end{equation}
Letting $\phi(y,\tau)=\sqrt{w(t)}\phi'(y,\tau)\exp(-i\theta_{S}(x,\tau))$, with $\theta_{S}(y,\tau)$ defined as in equation \eqref{EGBT theta 0}, then $\phi(y,\tau)$ satisfies the following Schrodinger equation
\begin{equation}
\label{EGBT Schrodinger equation t}
i\hbar\frac{\partial\phi}{\partial\tau}=-\frac{\hbar^{2}}{2m}\frac{\partial^{2}\phi}{\partial y^{2}}+mw(t)^{3}\left(\ddot{d(t)}y+\frac{\ddot{w(t)}}{2}y^{2}\right)\phi
\end{equation}
The potential term may be expressed in terms of $\tau$, similarly to before
\begin{equation}
\begin{multlined}
mw(t)^{3}\left(\ddot{d(t)}y+\frac{\ddot{w(t)}}{2}y^{2}\right) \\
=m\left(\left(\frac{\ddot{d(\tau)}}{w(\tau)}-2\left(\frac{\dot{d(\tau)}}{w(\tau)}\right)^{2}\right)y
+\left(\frac{\ddot{w(\tau)}}{2w(\tau)}-\left(\frac{\dot{w(\tau)}}{w(\tau)}\right)^{2}\right)y^{2}\right)
\end{multlined}
\end{equation}
We may relate $\phi(y,\tau)$ directly to the original wavefunction $\psi(x,t)$ by
\begin{equation}
\begin{aligned}
\phi(y,\tau)&=\sqrt{w(t)}\exp\left(-i\left(\theta_{S}(y,\tau)+\theta_{T}(y,\tau)\right)\right)\psi(y,\tau)\\
&=\sqrt{w(t)}\exp\left(-i\theta(y,\tau)\right)\psi(y,\tau)
\end{aligned}
\end{equation}
where
\begin{gather}
\label{EGBT theta}
\theta(y,t) = \frac{m}{2\hbar}\left(w(t)\dot{w(t)}y^{2}+2\dot{d(t)}w(t)y+\int_{0}^{t}\dot{d(t)}^{2}\,dt'\right) \\
\Leftrightarrow \theta(y,\tau)= \frac{m}{2\hbar}\left(\frac{\dot{w(\tau)}}{w(\tau)}y^{2}+2\frac{\dot{d(\tau)}}{w(\tau)}y+\int_{0}^{\tau}\left(\frac{\dot{d(\tau')}}{w(\tau')}\right)^{2}\,d\tau'\right) \\
\Leftrightarrow
\theta(x,t)=\frac{m}{2\hbar}\left(\frac{\dot{w(t)}}{w(t)}(x-d(t))^{2}+2\dot{d(t)}(x-d(t))+\int_{0}^{t}\dot{d(t)}^{2}\,dt'\right)
\end{gather}
$\theta$ may be expressed in terms of other coorindate systems but equation \eqref{EGBT theta} is it's simplest form.

The extended Greenberger transformation is a powerful technique; given a Schrodinger equation of the form \eqref{EGBT Schrodinger equation t} it may be possible to deduce a transformation that maps the problem to free space. In particular, we may generate solutions to \eqref{EGBT Schrodinger equation t} from free space solutions $\psi(x,t)$. However this mapping is not usually bijective. For $\phi(y,\tau)$ to satisfy equation \eqref{EGBT Schrodinger equation t} with $\ddot{d}$ and $\ddot{w}$ not vanishing, $\phi(y,\tau)$ should decay suitably as $|y|\rightarrow\infty$ to avoid infinite energies. This is not the case in free space \eqref{Schrodinger's equation}, where solutions of the form $\exp\left(ipx/\hbar\right)$ are allowed. For more information and applications, see \cite{Takagi:2}.

\subsection{The infinite potential well with slowly accelerating walls}

We are now in a position to study a wavefunction evolving in an infinite potential well with slowly accelerating walls. Consider such a wavefunction $\psi(x,t)$ evolving in a well with upper and lower walls $w_{2}(t)$ and $w_{1}(t)$ respectively. Then $\psi(x,t)$ must satisfy Schrodinger's equation \eqref{Schrodinger's equation} subject to the boundary conditions $\psi(w_{1}(t),t)=\psi(w_{1}(t),t)=0$. Let $w(t)=w_{2}(t)-w_{1}(t)$, and consider the extended Greenberger transformation induced by $y=(x-w_{1})/w$. By equations \eqref{EGBT Schrodinger equation t} and \eqref{EGBT theta}, define $\phi(y,\tau)$ as
\begin{equation}
\label{EGBT 0}
\begin{multlined}
\phi(y,\tau)=\sqrt{w(t)}\exp\left(-i\theta(y,\tau)\right)\psi(y,\tau) \\
=\sqrt{w(t)}\exp\left(-i\frac{m}{2\hbar}\left(w(t)\dot{w(t)}y^{2}+2\dot{w_{1}(t)}w(t)y+\int_{0}^{t}\dot{w_{1}(t)}^{2}\,dt'\right)\right)\psi(y,\tau)
\end{multlined}
\end{equation}
such that $\phi(y,\tau)$ satisfies
\begin{equation}
\label{General box SE}
\begin{aligned}
i\hbar\frac{\partial\phi}{\partial\tau}
&=-\frac{\hbar^{2}}{2m}\frac{\partial^{2}\phi}{\partial y^{2}}+
mw(t)^{3}\left(\ddot{w_{1}(t)}y+\frac{\ddot{w(t)}}{2}y^{2}\right)\phi \\
&=-\frac{\hbar^{2}}{2m}\frac{\partial^{2}\phi}{\partial y^{2}}+
\left(f(\tau)y+\frac{k(\tau)}{2}y^{2}\right)\phi
\end{aligned}
\end{equation}
subject to the boundary conditions $\phi(0,\tau)=\phi(1,\tau)=0$. The force and spring terms $f(t)=mw^{3}\ddot{w_{1}}$ and $k(t)=mw^{3}\ddot{w}/2$ have been introduced for notational convenience. 

A more symmetric transformation to use is $y=(x-w_{c})/w$, where $w_{m}=(w_{1}+w_{2})/2$ is the midpoint of the two walls. While this symmetry is elegant, the argument of the eigenmodes $\phi_{n}(y,\tau)$ simplify slightly in the coordinates we have chosen.

It is not possible to solve equation \eqref{General box SE} in general. However we may treat the fictitious terms $f(\tau)$ and $k(\tau)$ as a perturbation of the Hamiltonian. Explicitly,
\begin{equation}
\label{EGBT Hamiltonians}
i\hbar\frac{\partial \phi}{\partial \tau} 
= H\psi =\left(H_{f}+\Delta V\right)\psi 
= H_{f}\psi + \Delta V \psi
\end{equation}
where
\begin{equation}
H_{f}=-\frac{\hbar^{2}}{2m}\frac{\partial^{2}}{\partial y^{2}} \qquad \qquad \Delta V = f(\tau)y+\frac{k(\tau)}{2}y^{2}
\end{equation}
$\Delta V$ may be treated as a perturbation to $H_{f}$ if $\Delta V \ll H_{f}$. To make this comparison, we let $y=1$ and replace $H_{f}$ by it smallest eigenvalue, $\pi^{2}\hbar^{2}/2m$. 
\begin{equation}
\Delta V \ll H_{f} \Leftrightarrow \left|mw^{3}\left(\ddot{w_{1}}+\frac{\ddot{w}}{2}\right)\right| \ll \frac{\pi^{2}\hbar^{2}}{2m}
\end{equation}
For $\left|\ddot{w_{1}}+\ddot{w}/2\right|$ to be small, it is sufficient for $\left|\ddot{w_{i}}\right|$ to be small, where $i=1$ or $2$. Thus
\begin{equation}
\label{EGBT perturbation condition}
mw^{3}\left|\ddot{w_{i}}\right|\ll \frac{\pi^{2}\hbar^{2}}{2m} \Leftrightarrow \left|\ddot{w_{i}}\right| \ll \frac{1}{2w^{3}}\left(\frac{\pi\hbar}{m}\right)^{2}
\end{equation}
For an electron, $(\pi \hbar / m)\approx \textrm{cm}^{2}\textrm{ s}^{-1}$. In this case
\begin{equation}
\left|\ddot{w_{i}}\right| \ll \frac{\textrm{cm}^{4}\textrm{ s}^{-2}}{w^{3}}
\end{equation}
By equation $\eqref{EGBT perturbation condition}$, for $\Delta V \ll H_{f}$ both the width of the well $w(t)$ and the acceleration of the walls $\ddot{w_{i}(t)}$ must be small. In this case, we say that the well is slowly accelerating. Interestingly, no constraint is placed on the velocity of the walls. 

If both $|\ddot{w_{i}}|$ are small enough, we may ignore $\Delta V$ in comparison to $H_{f}$. This is the case we will study here, but an approximate solution is derived in appendix \ref{Appendix WKB}. In this case, equation \eqref{EGBT Hamiltonians} reduces to the free space Schrodinger equation and the problem reduces to an infinite potential well with constant width $1$. By the same logic as in section \ref{GB section}, we conclude that if the acceleration of each of the walls is slow enough, then
\begin{equation}
\label{Acceleration solutions}
\psi_{n}(x,t)=\sqrt{\frac{2}{w(t)}}\sin\left(\frac{n\pi(x-w_{1}(t))}{w(t)}\right)\exp\left(i\theta(y,t)-\frac{i}{\hbar}\int_{0}^{t}E_{n}(t')\,dt'\right)
\end{equation}
is a set of eigenmodes for the well where $\theta(y,t)$ is given in equation \eqref{EGBT 0}. Thus the evolution of a wavefunction in a well with slowly accelerating walls is very similar to the non-accelerating case; the only significant differences are the phase $\theta$ and the time variable $\tau$. In particular, equation \eqref{Psi propagation y} for the wavefunction revivals still holds true provdied we use definitions \eqref{EGBT theta} and \eqref{EGBT tau} for $\theta$ and $\tau$ is accordingly. In particular, we may still use the method of figure \ref{fig:Greenberger commutative diagram} to deduce the time evolution for $\psi(x,t)$.

\begin{figure}[t]
    \includegraphics[width=\textwidth]{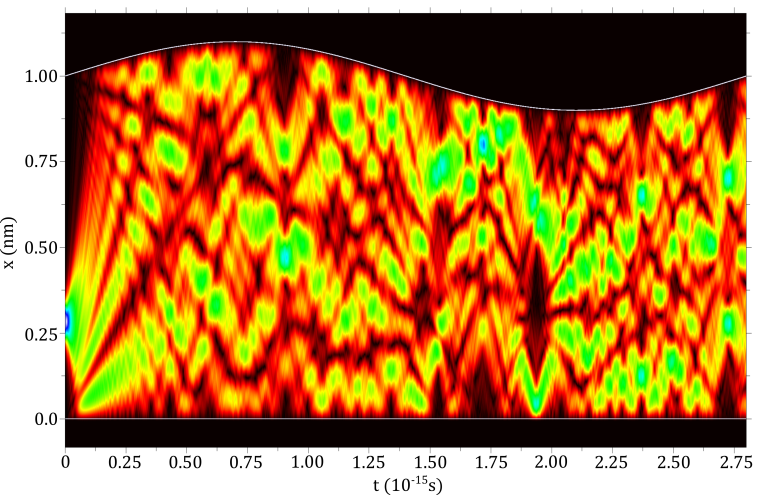}
    \caption{A wavefunction evolving in a well with one wall moving sinusoidally.}
    \centering
    \label{fig:sinusoidal simulation}
\end{figure}

Figure \ref{fig:sinusoidal simulation} is a simulation of $|\psi(x,t)|^{2}$ evolving in a well with one wall fixed and the other varying sinusoidally. The relative input location is the same as for figures \ref{fig:straight simulation} and \ref{fig:linear simulation}. As expected, the resulting evolution pattern is almost a stretched version of the previous two simulations. Some of the revivals appear less clear here than in the previous figures, such as the double revival in all three. This is because the eigenmodes \eqref{Acceleration solutions} are only approximate. The solutions would be more accurate if the acceleration of the wall was slower.

Note that there are no acceleration terms in equation \eqref{EGBT 0} for $\theta(y,t)$. In this sense the assumption that for small $|\ddot{w_{i}}|$ the accelerations may be ignored is consistent. The perturbation of these small accelerations to $w_{i}(t)$ and $\psi(x,t)$ is regular, as opposed to singular \cite{Bender}. $\theta(y,t)$ may be expressed in terms of $x$:
\begin{equation}
\label{Acceleration theta x}
\begin{aligned}
    \theta(x,t)&=\frac{m}{2\hbar}\left(\frac{\dot{w}}{w}(x-w_{1})^{2}+2\dot{w_{1}}x-2w_{1}\dot{w_{1}}+\int_{0}^{t}\dot{w_{1}}^{2}\,dt'\right) \\
    &=\frac{m}{2\hbar}\left(\frac{\dot{w}}{w}(x-w_{1})^{2}+2\dot{w_{1}}x-\int_{0}^{t}\dot{w_{1}}^{2}+w_{1}\ddot{w_{1}}\,dt'\right) 
\end{aligned}
\end{equation}
Equation \eqref{Acceleration theta x} is very similar to equation \eqref{Boosted Doescher Rice algebra}, where the variables $\Delta v$ and $v_{1}t$ have been replaced with $\dot{w}$ and $w_{1}(t)$ respecitvely, and the extra integration of $w_{1}\ddot{w_{1}}$ is required. 

When $\dot{w(t)}=0$, i.e. when the width of the well is locally constant for a given $t$, $\theta(y,t)$ simplifies to
\begin{gather}
\theta(y,t)=\frac{m}{\hbar}\left(\dot{w_{1}(t)}w(t)y+\frac{1}{2}\int_{0}^{t}\dot{w_{1}(t)}^{2}\,dt'\right) \\
\label{Acceleration theta fixed width}
\Rightarrow \theta(x,t)=\frac{m}{\hbar}\left(\dot{w_{1}(t)}x-\frac{1}{2}\int_{0}^{t}\dot{w_{1}(t)}^{2}\,dt'\right)
\end{gather}
where $\theta(x,t)$ has been derived at the end of appendix \ref{Galilean group appendix}, and we have neglected the acceleration term thereof. Equation \eqref{Acceleration theta fixed width} is an extension of equation \eqref{The parallel solution set} to time dependent velocities $v$. 

\begin{figure}[h]
    \def\svgwidth{\columnwidth}
    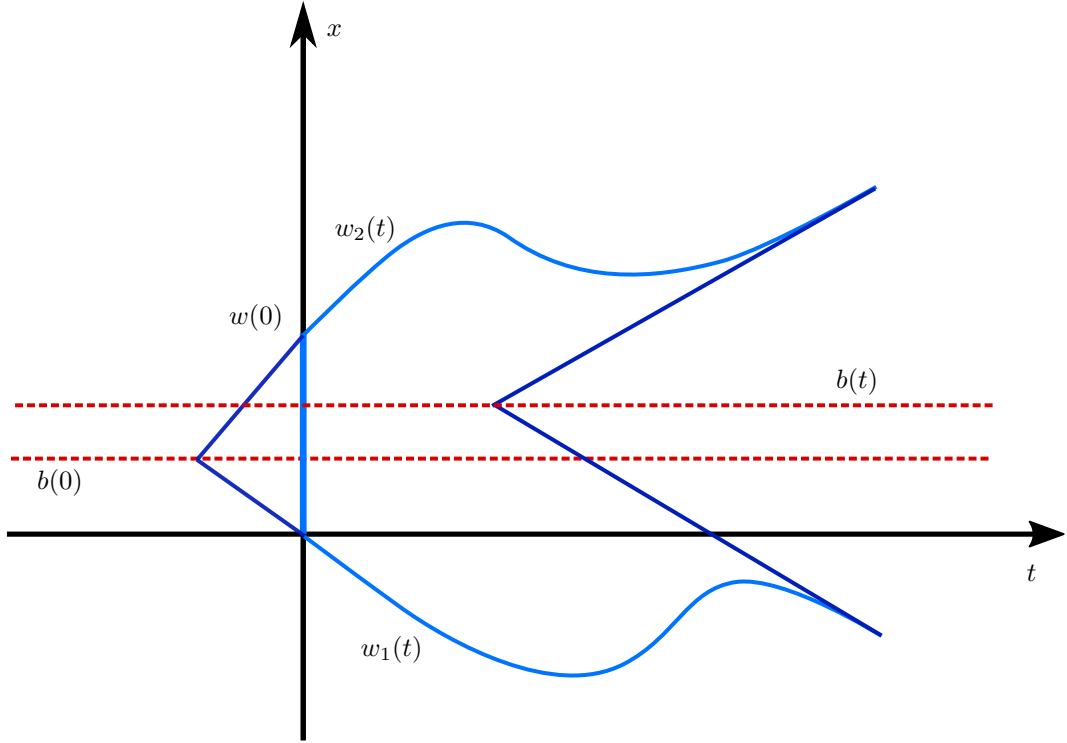
    \caption{An infinite potential well with slowly accelerating walls.}
    \label{fig:2}
    \centering
\end{figure}

Conversely, suppose that $\dot{w(t)}\neq0$. This allows us to complete the square in $y$:
\begin{gather}
\begin{aligned}
\theta(y,t)&=\frac{m}{2\hbar}\left(w\dot{w}y^{2}+2\dot{w_{1}}wy+\int_{0}^{t}\dot{w_{1}}^{2}\,dt'\right) \\
&=\frac{m}{2\hbar}\left(w\dot{w}\left(y+\frac{\dot{w_{1}}}{\dot{w}}\right)^{2}-\frac{w\dot{w_{1}}^{2}}{\dot{w}}+\int_{0}^{t}\dot{w_{1}}^{2}\,dt'\right) \\
&=\frac{m}{2\hbar}\left(w\dot{w}\left(y-c(t)\right)^{2}-\frac{w\dot{w_{1}}^{2}}{\dot{w}}+\int_{0}^{t}\dot{w_{1}}^{2}\,dt'\right)
\end{aligned} \\
\Leftrightarrow \theta(x,t)=\frac{m}{2\hbar}\left(\frac{\dot{w}}{w}\left(x-w_{1}(t)-b(t)\right)^{2}-\frac{w\dot{w_{1}}^{2}}{\dot{w}}+\int_{0}^{t}\dot{w_{1}}^{2}\,dt'\right)
\end{gather}
where 
\begin{equation}
    b(t) = -\frac{\dot{w_{1}}w}{\dot{w}}+w_{1} \qquad \qquad c(t) = -\frac{\dot{w_{1}}}{\dot{w}}
\end{equation}
$b(t)$ is just the instantaneous intersection of the walls, as illustrated in figure \ref{fig:2}, while $c(t)$ is this point in the $y$ coordinate. Again, this is just an extension of equation \eqref{The non parallel solution set}.

\begin{figure}[h]
    \def\svgwidth{\columnwidth}
    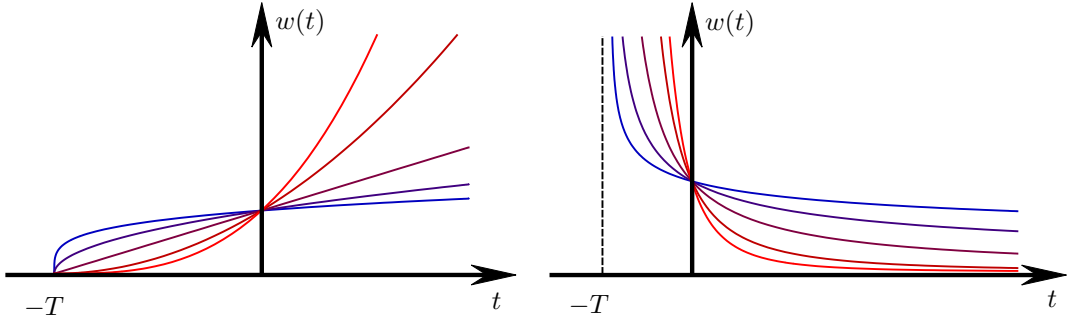
    \caption{Some different well shapes for $n>0$ (left) and $n<0$ (right), and fixed $T$ and $w_{0}$.}
    \label{fig:Monomial wells}
    \centering
\end{figure}

To investigate how $\tau$ and $\tau '$ behave for wells with slowly accelerating walls, we will consider widths $w(t)$ with a monomial like width dependency: $w(t)=w_{0}(1+(t/T))^{n}$ where $T\neq0$ is some fixed time and $n\in\mathbb{R}$. We will assume the well is symmetric about $x=0$, so that $w_{1}(t)=-w_{2}(t)$. For $T>0$, the well is expanding for $n>0$, fixed for $n=0$ and contracting for $n<0$. If $T<0$,  the well contracts to a point at $t=-T$ for $n>0$, is fixed for $n=0$ and expands to infinity at $t=-T$ for $n<0$. In the following, we will assume that $T>0$ and treat the negative $t$ values as evolving towards $-T$. This is equivalent to taking $T<0$ and evolving $t$ in the positive sense, however this way saves the amount of casework to be done. These different cases are illustrated in figure \ref{fig:Monomial wells}.

To ensure the walls are slowly accelerating, $w(t)$ must satisfy equation \eqref{EGBT perturbation condition}. As the well is symmetric, $|\ddot{w_{i}}|$ may be identified with $|\ddot{w}|$. Thus
\begin{equation}
\label{Slow monomial condition}
\frac{w_{0}^{4}n(n-1)}{T^{2}}\left(1+\frac{t}{T}\right)^{4n-2} \ll \left(\frac{\hbar}{m}\right)^{2}
\end{equation}
Equation \eqref{Slow monomial condition} is satisfied for all $t$ if $n=0$ or $1$, corresponding to the constant or linear width wells. Thus in the following, we will discard these two values of $n$. 

Let $\tau '$ be defined as
\begin{equation}
\tau '(t) = \frac{\hbar \pi}{2m}\tau(t) = \frac{\hbar \pi}{2m}\int_{0}^{t}\frac{dt'}{w(t')^{2}}
\end{equation}
$\tau'$ is a unit-less number, which for rational values describes the number of revivals at a particular time. For the present choice of $w(t)$:
\begin{gather}
\label{Monomial tau'}
\begin{aligned}
\tau'(t)&=\frac{\pi\hbar}{2mw_{0}^{2}}\int_{0}^{t}\frac{dt'}{(1+(t'/T))^{2n}} \\
&=\frac{\pi\hbar T}{2mw_{0}^{2}(1-2n)}\left(\frac{1}{(1+(t'/T))^{2n-1}}\right)_{0}^{t} \\
&=\sigma\left(\frac{1}{(1+(t/T))^{2n-1}}-1\right)
\end{aligned} \\
\Leftrightarrow
t(\tau')=T\left(\left(1+\frac{\tau'}{\sigma}\right)^{1/1-2n}-1\right)
\end{gather}
provided $n\neq 1/2$, where $\sigma = \pi\hbar T/2mw_{0}^{2}(1-2n)$.
For $n=1/2$,
\begin{gather}
\label{tau' 1/2}
\tau' = \sigma\log\left(1+\frac{t}{T}\right) \\
\Leftrightarrow t = T \left(e^{\tau'/\sigma}-1\right)
\end{gather}
where $\sigma=\pi\hbar T/2mw_{0}^{2}$. Note that both equations \eqref{Slow monomial condition} and \eqref{Monomial tau'} depend principally on $(1+t/T)^{2n-1}$. The limiting behaviour of $\tau'(t)$ is
\begin{gather}
\label{tau' inf}
\lim_{t\to\infty}\tau'(t)= \begin{cases}
\infty &\text{if } n \leq 1/2 \\
-\sigma &\text{if } n > 1/2 
\end{cases} \\
\label{tau' T}
\lim_{t\to -T}\tau'(t)= \begin{cases}
-\sigma &\text{if } n < 1/2 \\
\infty &\text{if } n \geq 1/2
\end{cases}
\end{gather}
We will separate our analysis into four cases: $n>1/2$, $n=1/2$, $0<n<1/2$ and $n<0$. 

For $n>1/2$, $(1+t/T)^{4n-2}\to0$ as $t\to -T$ but diverges as $t\to\infty$ (provided $n\neq 1$). Thus as $t$ tends to $-T$, the inequality \eqref{Slow monomial condition} is better satisfied and the slowly accelerating approximation becomes more valid as the width converges to $0$. By equation \eqref{tau' T}, $\tau'$ is unbounded as $t\to -T$. This is as expected, the frequency of the revivals should increase indefinitely as $w(t)\to0$.  However as $t$ gets larger, the slowly accelerating approximation becomes less valid. In particular, as $t$ approaches the order of
\begin{equation}
t\approx T \sqrt[2n-1]{\frac{T}{w_{0}^{2}}\frac{\hbar}{m}\frac{1}{\sqrt{n(n-1)}}}
\end{equation}
the approximation breaks down. For sufficiently small $t$, equation \eqref{tau' inf} holds true. To analyse $\phi(y,\tau)$ for larger values of $t$, we should solve equation \eqref{General box SE} directly or employ approximation methods as in appendix \ref{Appendix WKB}. 

If $n=1/2$ then the left hand side of equation $\eqref{Slow monomial condition}$ is $t$ independent, and the inequality is satisfied for all $t$ provided $T/w_{0}^{2}\gg m/\hbar$. $\tau '$ is unbounded, and attains all values from $-\infty$ to $\infty$ as $t$ varies from $-T$ to $\infty$ by equation \eqref{tau' 1/2}. The shape of the well appears as a horizontal parabola. 

If $0<n<1/2$, $(1+t/T)^{4n-2}\to 0$ and $\tau'\to\infty$ as $t\to \infty$ but $(1+t/T)^{4n-2}\to 0$ and $\tau'\to -\sigma$ as $t\to -T$ (provided $n \neq 0$). This is in contrast to the $n>1/2$ case, even though all $n>0$ $w(t)\to 0$ as $t\to -T$ and $w(t)$ diverges as $t\to \infty$. As $t\to \infty$, $w(t)$ accelerates very gently, but as $t\to -T$ $w(t)$ accelerates suddenly to $0$. As $w(t)$ grows, all possible revival times $\tau '$ are achieved. While the slowly accelerating approximation remains valid, $\tau'$ is bounded as $t\to -T$.

Finally, take $n<0$. In this case, $w(t)$ diverges as $t\to-T$ but converges smoothly to $0$ as $t\to\infty$. Thus it is not surprising that \eqref{Slow monomial condition} is satisfied and $\tau'$ is unbounded as $t\to\infty$. Conversely as $t\to-T$, \eqref{Slow monomial condition} becomes invalid but $\tau'$ is unbounded. 

\section{Conclusion}

After thoroughly expanding upon the particle in a box example, we now see that provided the acceleration of the walls is slow enough every well may be transformed into a well with fixed walls. Although a useful result in itself, many useful techniques for solving more general quantum mechanical problems have been developed along the way, particularly transformations methods. A wavfunction in a slowly accelerating well is seen to undergo revivals, and the location and phase of these revivals have been derived. For certain well geometries, we have deduced that certain revivals which occur in the fixed case may not occur in an expanding well, depending on the rate of expansion.

\appendix

\section{Symmetries of the free space Schrodinger equation}
\label{Niederer appendix}
In this appendix we will address certain symmetries of the free space Schrodinger equation as derived by Niederer \cite{Niederer}, and their relation to the Greenberger transformation \eqref{Greenberger 4}.

Interpreting \eqref{Greenberger 4} as a transformation of variables $(x,t)\mapsto(y,\tau)$ is convenient, but this transformation does not respect units. The problem is $w(t)$ which has units of distance, but ideally should be unit-less. This suggests substituting $1+\alpha t$ for $w(t)=w_{0}+\Delta v\,t$, where $\alpha$ has units of $\left(\textrm{time}\right)^{-1}$. In this form, we may interpret the transformation  $(x,t)\mapsto(y,\tau)$ as
\begin{equation}
\label{expansion transformation}
\left[\alpha\right](x,t)\mapsto\left(\frac{t}{1+\alpha t},\frac{x}{1+\alpha t}\right)
\end{equation}
Niederer \cite{Niederer} defined the transformation $\left[\alpha\right]$ as an \emph{expansion}. Expansions form an additive group on composition, $\left[\alpha_{1}\right]\circ\left[\alpha_{2}\right]=\left[\alpha_{1}+\alpha_{2}\right]$. Niederer discovered expansions while determining ``the maximal kinematical invariance group of the free Shcrodinger equation", i.e. the group of transformations $(x,t)\mapsto g(x,t)$ such that the map $\psi(x,t)\mapsto f_{g}(g(x,t))\psi(g(x,t))$ preserves solutions of Schrodinger's equation, where $f_{g}(x,t)$ is some adjustment factor. He determined that all such symmetries (which we are restricting to 1 spatial dimension) are of the form
\begin{equation}
\label{Niederer transformation}
g(x,t)=\left(d^{2}\frac{t+b}{1+\alpha(t+b)},d\frac{x+vt+a}{1+\alpha(t+b)}\right)
\end{equation}
for real parameters $d,\alpha,b,a$ and $v$. These parameters correspond respectively to dilations, expansions, temporal translations, spatial translations and Galilean boosts. For example, the transformation $(x,t)\mapsto(y,\tau)=g(x,t)$ corresponds to $\alpha = \Delta v/w_{0}$ and $d=1/w_{0}$. Note that we have already derived some of the functions $f_{g}(x,t)$, e.g. for Gallilean boosts in equation \eqref{moving reference frame transformation}. 

Of these 5 transformations, the expansion is (perhaps) the only counter intuitive one. The expansion may be derived from the \emph{Appell transformation} $\Sigma$:
\begin{equation}
\label{Appell transformation}
\Sigma(x,t)=\left(\frac{x}{t},-\frac{1}{t}\right)
\end{equation}
In 1892 Appell \cite{Appell} showed that the Appel transformation is a symmetry of the heat equation $u_{t}(x,t)-k\,u_{xx}(x,t)=0$ \cite{Widder}. As the Schrodinger equation is a complex variant of the heat equation with $k=-i\hbar/2m$, the Appell transformation is a symmetry of the Schrodinger equation also. 
Denote the time translation $(x,t)\mapsto(x,t+b)$ by $(b)$. Niederer showed that
\begin{equation}
\left[\alpha\right](x,t)=\Sigma^{-1}(-\alpha)\Sigma(x,t)=\left(\frac{t}{1+\alpha t},\frac{x}{1+\alpha t}\right)
\end{equation}
Thus Greenberger's method is a result of temporal symmetry and Appell's transformation. As Niederer's transformations \eqref{Niederer transformation} form a closed group, there is nothing to be gained from applying Greenberger's transformation iteratively; the result must again be of the form \eqref{Niederer transformation}.

Note that Niederer's transformation \eqref{Niederer transformation} also respects boundary conditions nicely like Greenberger's transformation, as the transformed wavefunction is just multiplied by $f_{g}$. 

\section{The group structure of the extended Galilean transformations}
\label{Galilean group appendix}

The extended Galilean transformations form a group, as translating by $d_{1}(t)$ and then by $d_{2}(t)$ is equivalent to translating by $(d_{1}+d_{2})(t)$. However as we have defined it, the transformation $\psi(x,t)\mapsto\phi(x',t)$ does not respect this group structure. 
To see this, consider two consecutive transformations
\begin{equation}
\label{2 EGTs}
(x,t)\mapsto(x_{1},t)=(x-d_{1}(t),t) \qquad \quad (x_{1},t)\mapsto(x_{2},t)=(x_{1}-d_{2}(t),t)
\end{equation}
If the potential in the $(x,t)$ frame is $V$, then by equation \eqref{EGT V transformation}, $V_{1}=V(x-d_{1}(t))+m\ddot{d_{1}}x_{1}$ and
\begin{equation}
\begin{aligned}
V_{2}&=V(x-d_{1}(t)-d_{2}(t))+m\ddot{d_{1}}\left(x_{2}+d_{2}(t)\right)+m\ddot{d_{2}}x_{2} \\
&=V(x-(d_{1}-d_{2})(t))+m(\ddot{d_{1}+d_{2}})x_{2}+m\ddot{d_{1}}d_{2}
\end{aligned}
\end{equation}
$V_{2}$ differs from the induced potential of the transformation $x\mapsto x-\left(d_{1}+d_{2}\right)(t)$ by $m\ddot{d_{1}}d_{2}$. There is no spatial dependence in $m\ddot{d_{1}}d_{2}$, and so it produces no force. This term may be integrated out by multiplying $\phi$ by $\exp\left(-i\hbar\int m\ddot{d_{2}}d_{1}\,dt'\right)$.

Thus for the product of two Galillean transformations as in equation \eqref{2 EGTs}, let
\begin{equation}
\label{EGT group 1}
\phi_{1}(x_{1},t)=\exp\left(-i\theta_{1}(x_{1},t)\right)\psi(x_{1},t) \qquad 
\phi_{2}(x_{2},t)=\exp\left(-i\theta_{2}(x_{2},t)\right)\phi_{1}(x_{2},t)
\end{equation}
where
\begin{gather}
\label{EGT group 2}
\theta_{1}(x_{1},t)=\frac{m}{\hbar}\left(\dot{d_{1}}x_{1}+\int_{0}^{t}\frac{\dot{d_{1}}^{2}}{2}\,dt'\right) \\
\label{EGT theta non inertial}
\theta(x_{2},t)=\frac{m}{\hbar}\left(\dot{d_{2}}x_{2}+\int_{0}^{t}\frac{\dot{d_{2}}^{2}}{2}-\ddot{d_{1}}d_{2}\,dt'\right)
\end{gather}
Then $\phi_{2}(x_{2},t)=\exp(-i(\theta_{2}(x_{2},t)+\theta_{1}(x_{2},t)))\psi(x_{2},t)$, where
\begin{equation}
\begin{aligned}
\theta_{2}(x_{2},t)+\theta_{1}(x_{2},t) & =\frac{m}{\hbar}\left(\dot{d_{2}}x_{2}+\dot{d_{1}}(x_{1}+d_{2})+\frac{1}{2}\int_{0}^{t}\dot{d_{1}}^{2}+\dot{d_{2}}^{2}-2\ddot{d_{1}}d_{2}\,dt'\right) \\
&=\frac{m}{\hbar}\left((\dot{d_{1}+d_{2}})x_{2}+\dot{d_{1}}d_{2}+\frac{1}{2}\int_{0}^{t}\dot{d_{1}}^{2}+\dot{d_{2}}^{2}-2\ddot{d_{1}}d_{2}\,dt'\right)
\end{aligned}
\end{equation}
By integration by parts:
\begin{equation}
\int_{0}^{t}\ddot{d_{1}}d_{2}\,dt'=\left(\dot{d_{1}}d_{2}\right)_{0}^{t}-\int_{0}^{t}\dot{d_{1}}\dot{d_{2}}\,dt'
\end{equation}
Thus up to a constant $\dot{d_{1}}(0)d_{2}(0)$,
\begin{equation}
\theta_{2}(x_{2},t)+\theta_{1}(x_{2},t)=\frac{m}{\hbar}\left((\dot{d_{1}+d_{2}})x_{2}+\frac{1}{2}\int_{0}^{t}\dot{(d_{1}+d_{2})}^{2}\,dt'\right)
\end{equation}
which is exactly the phase change required for a translation by $(d_{1}+d_{2})(t)$ by equation \eqref{EGT 5}. Further, $\phi_{2}(x_{2},t)$ perceives a potential $V=m\ddot{(d_{1}+d_{2})}x_{2}$.
Thus the transformations defined by equations \eqref{EGT group 1}, \eqref{EGT group 2} and \eqref{EGT theta non inertial} respect the group structure of the extended Galilean transformations. The disadvantage of this definition is that it must be defined with respect to some underlying reference frame $(x,t)$. Note that equation \eqref{EGT theta non inertial} correctly reduces to equation \eqref{EGT group 2} when $\ddot{d_{1}}=0$. 

If $\psi(x,t)\mapsto\phi(x',t)=\psi(x',t)\exp\left(-i\theta(x',t)\right)$ is induced by a translation $x'=x-d(t)$, then we may express $\theta$ in terms of the original coordinate $x$:
\begin{equation}
\label{EGT theta inversion}
\begin{aligned}
\theta(x,t)&=\frac{m}{\hbar}\left(\dot{d}(x+d(t))+\frac{1}{2}\int_{0}^{t}\dot{d}^{2}\,dt'\right) \\
&=\frac{m}{\hbar}\left(\dot{d}x+d\dot{d}+\frac{1}{2}\int_{0}^{t}\dot{d}^{2}\,dt'\right) \\
&=\frac{m}{\hbar}\left(+\dot{d}x-\int_{0}^{t}\frac{\dot{d}^{2}}{2}+d\ddot{d}\,dt'\right)
\end{aligned}
\end{equation}
up to a constant. Here we have differentiated and then integrated $d\dot{d}$ to bring this term inside the integral. Equation \eqref{EGT theta inversion} allows us to invert an extended Galilean transformation via $\psi(x,t)=\phi(x,t)\exp\left(i\theta(x,t)\right)$. When $\ddot{d}=0$, $\dot{d}$ is constant and the integral is just $\dot{d}^{2}t/2$. This result agrees with equation \eqref{The parallel solution set} where $\dot{d}=v$. Equation \eqref{EGT theta inversion} can also be deduced by considering the product of the translations by $d(t)$ and $-d(t)$ and applying equation \eqref{EGT theta non inertial}.

\section{The WKB approximation}

\label{Appendix WKB}
In this appendix we will provide an approximate solution to equation \eqref{General box SE} using the adiabatic and WKB approximations. 
\begin{equation}
\begin{aligned}
i\hbar\frac{\partial\phi}{\partial\tau}
&=-\frac{\hbar^{2}}{2m}\frac{\partial^{2}\phi}{\partial y^{2}}+
\left(f(\tau)y+\frac{k(\tau)}{2}y^{2}\right)\phi \\
&=H(\tau)\psi = H_{f}\psi+\Delta V(\tau)\psi
\end{aligned}
\end{equation}
subject to the boundary conditions $\phi(0,\tau)=\phi(1,\tau)=0$.

The first assumption we make is that $H(\tau)$ varies slow enough so that the adiabatic approximation may be used, as in equation \eqref{Adiabatic solutions}. As $H_{f}$ is time independent, this is equivalent to requiring that $\Delta V(\tau)$ be slowly varying. Thus, we will solve for the instantaneous eigenmodes $\phi_{n}(y,\tau)$:
\begin{equation}
    E_{n}(\tau)\psi=H(\tau)\psi_{n}(y,\tau)
\end{equation}
where $E_{n}(\tau)$ is the associated instantaneous eigenvalue.

We will further assume that $\Delta V $ is slowly varying in time, but we will not derive the necessary conditions on $w_{i}$ for this to be so. This allows us to make the adiabatic approximation in solving for for $\phi(y,\tau)$; i.e. we approximate the instantaneous eigenmodes $\phi_{n}(y,\tau)$ of the Hamiltonian $H(\tau)$ along with a dynamical phase and Berry phase as solutions to equation \eqref{EGBT Hamiltonians}. The instantaneous eigenmodes and eigenvalues $E_{n}(\tau)$ are given by
\begin{equation}
\label{EGBT instant modes}
E_{n}(\tau)\phi_{n}(y,\tau)=H(\tau)\phi_{n}(y,\tau)
\end{equation}

To solve equation \eqref{EGBT instant modes}, we shall employ the \emph{WKB approximation}. Consider the one dimensional time independent Schrodinger equation
\begin{equation}
\label{WKB 1}
\begin{gathered}
-\frac{\hbar^{2}}{2m}\frac{d^{2}\psi}{dx^{2}}+V(x)\psi=E\psi \\
\Leftrightarrow \frac{d^{2}\psi}{dx^{2}}=\frac{2m}{\hbar^{2}}(V(x)-E)\psi
\end{gathered}
\end{equation}
The idea of the WKB approximation is to replace $\psi(x)$ by an exponential function $\exp(i\Theta(x))$. Upon substitution into equation \eqref{WKB 1}:
\begin{equation}
\label{WKB 2}
i\frac{d^{2}\Theta}{dx^{2}}-\left(\frac{d\Theta}{dx}\right)^{2}=\frac{2m}{\hbar^{2}}(V(x)-E)
\end{equation}
To first order, the WKB approximation neglects derivatives of $\Theta(x)$ higher than the first. Thus we may solve equation \eqref{WKB 2}:
\begin{equation}
\label{WKB 3}
\Theta(x)=\pm\int_{0}^{x}\sqrt{\frac{2m}{\hbar^{2}}\left(E-V(x')\right)}\,dx'+C
\end{equation}
where $C$ is some constant. For a more thorough discussion of the WKB approximation and extension to higher orders, see \cite{Bender,Shankar}.

By equation \eqref{WKB 3} and the linearity of \eqref{EGBT instant modes}, $\phi_{n}(y,\tau)$ will have the form
\begin{equation}
\begin{multlined}
\phi_{n}(y,\tau)=A\exp\left(\int_{0}^{y}\sqrt{\frac{2m}{\hbar^{2}}\left(E_{n}(\tau)-\Delta V(y')\right)}\,dy'+C\right) \\+B\exp\left(\int_{0}^{y}\sqrt{\frac{2m}{\hbar^{2}}\left(E_{n}(\tau)-\Delta V(y')\right)}\,dy'+C\right)
\end{multlined}
\end{equation}
where $A$ and $B$ are complex constants. To satisfy $\phi_{n}(0,\tau)=0$, we further set
\begin{equation}
\label{WKB wavefunction}
\phi_{n}(y,\tau)=\sqrt{2}\sin\left(\int_{0}^{y}\sqrt{\frac{2m}{\hbar^{2}}\left(E_{n}(\tau)-\Delta V(y')\right)}\,dy'\right)
\end{equation}
Note that we have ignored the effect of the perturbation $\Delta V$ on the normalisation constant $\sqrt{2}$. 
To satisfy the other boundary condition, $\phi(1,\tau)=0$ we require
\begin{equation}
\label{WKB BC integral}
\int_{0}^{1}\sqrt{\frac{2m}{\hbar^{2}}\left(E_{n}(\tau)-\Delta V(y')\right)}\,dy' = n\pi
\end{equation}
for some integer $n$, which coincides with the mode number. 

To evaluate the integral \eqref{WKB BC integral} we will perform a binomial expansion; $\sqrt{1+x}\approx1+x/2$ for $|x|\ll1$.
\begin{equation}
\begin{aligned}
\int_{0}^{1}\sqrt{\frac{2m}{\hbar^{2}}\left(E_{n}(\tau)-\Delta V(y')\right)}\,dy' &=\sqrt{\frac{2mE_{n}}{\hbar^{2}}}\int_{0}^{1}\sqrt{1-\frac{\Delta V }{E_{n}}}\,dy' \\
&\approx\sqrt{\frac{2mE_{n}}{\hbar^{2}}}\int_{0}^{1}1-\frac{\Delta V }{2E_{n}}\,dy'\\
&=\sqrt{\frac{2mE_{n}}{\hbar^{2}}}-\frac{1}{2}\sqrt{\frac{2m}{\hbar^{2}E_{n}}}\int_{0}^{1}\Delta V(y')\,dy'
\end{aligned}
\end{equation}
We will denote $\int_{0}^{1}\Delta V(y')\,dy'$ by $\int\Delta V$. By direct integration,
\begin{equation}
\int\Delta V(\tau) = \frac{f(\tau)}{2}+\frac{k(\tau)}{6}
\end{equation}
However the current argument works for any perturbing potential $\Delta V$ in an infinite potential well.
By equation \eqref{WKB BC integral},
\begin{equation}
\label{WKB energy}
\begin{gathered}
\sqrt{E_{n}}-\frac{\int\Delta V}{2\sqrt{E_{n}}} =n\pi\sqrt{\frac{\hbar^{2}}{2m}} \\
\Rightarrow E_{n}-n\pi\sqrt{\frac{\hbar^{2}}{2m}}\sqrt{E_{n}}-\textstyle{\int}\Delta V/2=0 \\
\begin{aligned}
\Rightarrow \sqrt{E_{n}}&=\frac{n\pi\sqrt{\hbar^{2}/2m}+\sqrt{(n\pi)^{2}\hbar^{2}/2m+2\textstyle{\int}\Delta V}}{2} \\
&=n\pi\sqrt{\frac{\hbar^{2}}{2m}}\left(\frac{1+\sqrt{1+(4m\textstyle{\int}\Delta V/(n\hbar\pi)^{2})}}{2}\right)
\end{aligned}
\end{gathered}
\end{equation}

\begin{equation}
\begin{gathered}
\Rightarrow \sqrt{E_{n}}\approx n\pi\sqrt{\frac{\hbar^{2}}{2m}}\left(1+\frac{m\textstyle{\int}\Delta V}{(n\hbar\pi)^{2}}\right) \\
\Rightarrow E_{n}(\tau)\approx n^{2}\pi^{2}\frac{\hbar^{2}}{2m}\left(1 + \frac{2m\textstyle{\int}\Delta V}{(n\hbar\pi)^{2}}\right)= \frac{\hbar^{2}n^{2}\pi^{2}}{2m}+\int\Delta V
\end{gathered}
\end{equation}
where we have twice used the binomial expansion $(1+x)^{n}\approx 1+nx$ for $|x|\ll 1$.  By equation \eqref{WKB energy}, to first order $\Delta V$ shifts all the energy levels $E_{n}(\tau)$ from the unperturbed energies by the same amount $\int\Delta V$. 

Equation \eqref{WKB energy} along with equation \eqref{WKB wavefunction} gives an approximate expression for $\phi_{n}(y,\tau)$:
\begin{equation}
\begin{aligned}
\phi_{n}(y,\tau)&\approx\sqrt{2}\sin\left(\int_{0}^{y}\sqrt{\frac{2m}{\hbar^{2}}\left(E_{n}(\tau)-\Delta V(y')\right)}\,dy'\right) \\
&\approx \sqrt{2}\sin\left(n\pi y +\frac{m\hbar^{2}}{n\pi}\left(\Delta Vy-\int_{0}^{y}\Delta V(y')\,dy'\right)\right)
\end{aligned}
\end{equation}
For $V(\tau)=f(\tau)y+k(\tau)y^{2}/2$,
\begin{equation}
\phi_{n}(y,\tau)\approx\sqrt{2}\sin\left(n\pi y +\frac{m\hbar^{2}}{n\pi}\left(\frac{f(\tau)}{2}y(1-y)+\frac{k(\tau)}{6}y(1-y^{2})\right)\right)
\end{equation}
Note that the effect of $\Delta V$ on the mode shape decreases with mode number. It is possible to calculate the Berry phase $\gamma$ of the time evolution using $\sin(x+\Delta x)\approx\sin(x)+\cos(x)\Delta x$, but we will not do so here. 

\bibliographystyle{unsrt}
\bibliography{references}
\end{document}

%% file: moving_walls_schematic_tex.pdf_tex
\begingroup%
  \makeatletter%
  \providecommand\color[2][]{%
    \errmessage{(Inkscape) Color is used for the text in Inkscape, but the package 'color.sty' is not loaded}%
    \renewcommand\color[2][]{}%
  }%
  \providecommand\transparent[1]{%
    \errmessage{(Inkscape) Transparency is used (non-zero) for the text in Inkscape, but the package 'transparent.sty' is not loaded}%
    \renewcommand\transparent[1]{}%
  }%
  \providecommand\rotatebox[2]{#2}%
  \ifx\svgwidth\undefined%
    \setlength{\unitlength}{650.51389679bp}%
    \ifx\svgscale\undefined%
      \relax%
    \else%
      \setlength{\unitlength}{\unitlength * \real{\svgscale}}%
    \fi%
  \else%
    \setlength{\unitlength}{\svgwidth}%
  \fi%
  \global\let\svgwidth\undefined%
  \global\let\svgscale\undefined%
  \makeatother%
  \begin{picture}(1,0.7068213)%
    \put(0,0){\includegraphics[width=\unitlength,page=1]{moving_walls_schematic_tex.pdf}}%
    \put(0.30806109,0.66335391){\color[rgb]{0,0,0}\makebox(0,0)[lb]{\smash{$x$}}}%
    \put(0.94756195,0.16129697){\color[rgb]{0,0,0}\makebox(0,0)[lb]{\smash{$t$}}}%
    \put(0.29566323,0.28945185){\color[rgb]{0,0,0}\makebox(0,0)[lb]{\smash{$b$}}}%
    \put(0.24567229,0.39994007){\color[rgb]{0,0,0}\makebox(0,0)[lb]{\smash{$w_{0}$}}}%
    \put(0.48396577,0.54287748){\color[rgb]{0,0,0}\makebox(0,0)[lb]{\smash{$w_{0}+v_{2}t$}}}%
    \put(0.52491117,0.09071163){\color[rgb]{0,0,0}\makebox(0,0)[lb]{\smash{$v_{1}t$}}}%
  \end{picture}%
\endgroup%

%% file: moving_reference_frame_tex.pdf_tex
\begingroup%
  \makeatletter%
  \providecommand\color[2][]{%
    \errmessage{(Inkscape) Color is used for the text in Inkscape, but the package 'color.sty' is not loaded}%
    \renewcommand\color[2][]{}%
  }%
  \providecommand\transparent[1]{%
    \errmessage{(Inkscape) Transparency is used (non-zero) for the text in Inkscape, but the package 'transparent.sty' is not loaded}%
    \renewcommand\transparent[1]{}%
  }%
  \providecommand\rotatebox[2]{#2}%
  \ifx\svgwidth\undefined%
    \setlength{\unitlength}{1427.88571335bp}%
    \ifx\svgscale\undefined%
      \relax%
    \else%
      \setlength{\unitlength}{\unitlength * \real{\svgscale}}%
    \fi%
  \else%
    \setlength{\unitlength}{\svgwidth}%
  \fi%
  \global\let\svgwidth\undefined%
  \global\let\svgscale\undefined%
  \makeatother%
  \begin{picture}(1,0.32047385)%
    \put(0,0){\includegraphics[width=\unitlength,page=1]{moving_reference_frame_tex.pdf}}%
    \put(0.14170667,0.29956847){\color[rgb]{0,0,0}\makebox(0,0)[lb]{\smash{$x$}}}%
    \put(0.43416999,0.0629984){\color[rgb]{0,0,0}\makebox(0,0)[lb]{\smash{$t$}}}%
    \put(0.09063527,0.16018944){\color[rgb]{0,0,0}\makebox(0,0)[lb]{\smash{$w_{0}$}}}%
    \put(0.19879398,0.25780831){\color[rgb]{0,0,0}\makebox(0,0)[lb]{\smash{$w_{0}+\Delta vt$}}}%
    \put(0,0){\includegraphics[width=\unitlength,page=2]{moving_reference_frame_tex.pdf}}%
    \put(0.67818014,0.29943727){\color[rgb]{0,0,0}\makebox(0,0)[lb]{\smash{$x$}}}%
    \put(0.97064353,0.06411894){\color[rgb]{0,0,0}\makebox(0,0)[lb]{\smash{$t$}}}%
    \put(0.6736525,0.12797503){\color[rgb]{0,0,0}\makebox(0,0)[lb]{\smash{$b$}}}%
    \put(0.6284669,0.16133235){\color[rgb]{0,0,0}\makebox(0,0)[lb]{\smash{$w_{0}$}}}%
    \put(0.74711318,0.20989091){\color[rgb]{0,0,0}\makebox(0,0)[lb]{\smash{$w_{0}+v_{2}t$}}}%
    \put(0.74760096,0.03855382){\color[rgb]{0,0,0}\makebox(0,0)[lb]{\smash{$v_{1}t$}}}%
    \put(0,0){\includegraphics[width=\unitlength,page=3]{moving_reference_frame_tex.pdf}}%
  \end{picture}%
\endgroup%

%% file: Greenberger_transformation_tex.pdf_tex
\begingroup%
  \makeatletter%
  \providecommand\color[2][]{%
    \errmessage{(Inkscape) Color is used for the text in Inkscape, but the package 'color.sty' is not loaded}%
    \renewcommand\color[2][]{}%
  }%
  \providecommand\transparent[1]{%
    \errmessage{(Inkscape) Transparency is used (non-zero) for the text in Inkscape, but the package 'transparent.sty' is not loaded}%
    \renewcommand\transparent[1]{}%
  }%
  \providecommand\rotatebox[2]{#2}%
  \ifx\svgwidth\undefined%
    \setlength{\unitlength}{825.59029166bp}%
    \ifx\svgscale\undefined%
      \relax%
    \else%
      \setlength{\unitlength}{\unitlength * \real{\svgscale}}%
    \fi%
  \else%
    \setlength{\unitlength}{\svgwidth}%
  \fi%
  \global\let\svgwidth\undefined%
  \global\let\svgscale\undefined%
  \makeatother%
  \begin{picture}(1,0.31827172)%
    \put(0,0){\includegraphics[width=\unitlength,page=1]{Greenberger_transformation_tex.pdf}}%
    \put(0.00459258,0.10461777){\color[rgb]{0,0,0}\makebox(0,0)[lb]{\smash{$x$}}}%
    \put(0.12945563,0.01796117){\color[rgb]{0,0,0}\makebox(0,0)[lb]{\smash{$t$}}}%
    \put(0.51942302,0.10738635){\color[rgb]{0,0,0}\makebox(0,0)[lb]{\smash{$y$}}}%
    \put(0.64650086,0.01796117){\color[rgb]{0,0,0}\makebox(0,0)[lb]{\smash{$\tau$}}}%
    \put(0,0){\includegraphics[width=\unitlength,page=2]{Greenberger_transformation_tex.pdf}}%
  \end{picture}%
\endgroup%

%% file: Double_revival_tex.pdf_tex
\begingroup%
  \makeatletter%
  \providecommand\color[2][]{%
    \errmessage{(Inkscape) Color is used for the text in Inkscape, but the package 'color.sty' is not loaded}%
    \renewcommand\color[2][]{}%
  }%
  \providecommand\transparent[1]{%
    \errmessage{(Inkscape) Transparency is used (non-zero) for the text in Inkscape, but the package 'transparent.sty' is not loaded}%
    \renewcommand\transparent[1]{}%
  }%
  \providecommand\rotatebox[2]{#2}%
  \ifx\svgwidth\undefined%
    \setlength{\unitlength}{549.60001788bp}%
    \ifx\svgscale\undefined%
      \relax%
    \else%
      \setlength{\unitlength}{\unitlength * \real{\svgscale}}%
    \fi%
  \else%
    \setlength{\unitlength}{\svgwidth}%
  \fi%
  \global\let\svgwidth\undefined%
  \global\let\svgscale\undefined%
  \makeatother%
  \begin{picture}(1,0.41193594)%
    \put(0,0){\includegraphics[width=\unitlength,page=1]{Double_revival_tex.pdf}}%
    \put(0.00607386,0.14159769){\color[rgb]{0,0,0}\makebox(0,0)[lb]{\smash{$y$}}}%
    \put(0.19405463,0.00726624){\color[rgb]{0,0,0}\makebox(0,0)[lb]{\smash{$\tau$}}}%
    \put(0.13828238,0.16011644){\color[rgb]{0,0,0}\makebox(0,0)[lb]{\smash{$\phi\left(y,\tau'=0\right)$}}}%
    \put(0.82203247,0.18723843){\color[rgb]{0,0,0}\makebox(0,0)[lb]{\smash{$\phi\left(y,\tau'=\frac{1}{2}\right)$}}}%
  \end{picture}%
\endgroup%

%% file: Greenberger_commutative_tex.pdf_tex
\begingroup%
  \makeatletter%
  \providecommand\color[2][]{%
    \errmessage{(Inkscape) Color is used for the text in Inkscape, but the package 'color.sty' is not loaded}%
    \renewcommand\color[2][]{}%
  }%
  \providecommand\transparent[1]{%
    \errmessage{(Inkscape) Transparency is used (non-zero) for the text in Inkscape, but the package 'transparent.sty' is not loaded}%
    \renewcommand\transparent[1]{}%
  }%
  \providecommand\rotatebox[2]{#2}%
  \ifx\svgwidth\undefined%
    \setlength{\unitlength}{825.31424935bp}%
    \ifx\svgscale\undefined%
      \relax%
    \else%
      \setlength{\unitlength}{\unitlength * \real{\svgscale}}%
    \fi%
  \else%
    \setlength{\unitlength}{\svgwidth}%
  \fi%
  \global\let\svgwidth\undefined%
  \global\let\svgscale\undefined%
  \makeatother%
  \begin{picture}(1,0.33947243)%
    \put(0.22452892,0.25481962){\color[rgb]{0,0,0}\makebox(0,0)[lb]{\smash{$\psi(x,0)$}}}%
    \put(0.21817228,0.05118205){\color[rgb]{0,0,0}\makebox(0,0)[lb]{\smash{$\phi(y,0)$}}}%
    \put(0.66003462,0.25481962){\color[rgb]{0,0,0}\makebox(0,0)[lb]{\smash{$\psi(x,t)$}}}%
    \put(0.66695835,0.05347643){\color[rgb]{0,0,0}\makebox(0,0)[lb]{\smash{$\phi(y,\tau)$}}}%
    \put(0.33409948,0.27636228){\color[rgb]{0,0,0}\makebox(0,0)[lb]{\smash{Time evolution in $t$}}}%
    \put(0.32992359,0.01837477){\color[rgb]{0,0,0}\makebox(0,0)[lb]{\smash{Time evolution in $\tau$}}}%
    \put(0,0){\includegraphics[width=\unitlength,page=1]{Greenberger_commutative_tex.pdf}}%
    \put(0.21510503,0.14841998){\color[rgb]{0,0,0}\makebox(0,0)[lb]{\smash{$G$}}}%
    \put(0.71018396,0.14871343){\color[rgb]{0,0,0}\makebox(0,0)[lb]{\smash{$G^{-1}$}}}%
    \put(0,0){\includegraphics[width=\unitlength,page=2]{Greenberger_commutative_tex.pdf}}%
  \end{picture}%
\endgroup%

%% file: Expanding_double_revival_tex.pdf_tex
\begingroup%
  \makeatletter%
  \providecommand\color[2][]{%
    \errmessage{(Inkscape) Color is used for the text in Inkscape, but the package 'color.sty' is not loaded}%
    \renewcommand\color[2][]{}%
  }%
  \providecommand\transparent[1]{%
    \errmessage{(Inkscape) Transparency is used (non-zero) for the text in Inkscape, but the package 'transparent.sty' is not loaded}%
    \renewcommand\transparent[1]{}%
  }%
  \providecommand\rotatebox[2]{#2}%
  \ifx\svgwidth\undefined%
    \setlength{\unitlength}{520.40618656bp}%
    \ifx\svgscale\undefined%
      \relax%
    \else%
      \setlength{\unitlength}{\unitlength * \real{\svgscale}}%
    \fi%
  \else%
    \setlength{\unitlength}{\svgwidth}%
  \fi%
  \global\let\svgwidth\undefined%
  \global\let\svgscale\undefined%
  \makeatother%
  \begin{picture}(1,0.64284049)%
    \put(0,0){\includegraphics[width=\unitlength,page=1]{Expanding_double_revival_tex.pdf}}%
    \put(0.02560604,0.16554039){\color[rgb]{0,0,0}\makebox(0,0)[lb]{\smash{$x$}}}%
    \put(0.19202979,0.03361576){\color[rgb]{0,0,0}\makebox(0,0)[lb]{\smash{$t$}}}%
    \put(0.11821052,0.21079656){\color[rgb]{0,0,0}\makebox(0,0)[lb]{\smash{$\psi\left(x,t=0\right)$}}}%
    \put(0.86859716,0.27833749){\color[rgb]{0,0,0}\makebox(0,0)[lb]{\smash{$\psi\left(x,t\left(\tau'=\frac{1}{2}\right)\right)$}}}%
    \put(0,0){\includegraphics[width=\unitlength,page=2]{Expanding_double_revival_tex.pdf}}%
  \end{picture}%
\endgroup%

%% file: t_of_tau_tex.pdf_tex
\begingroup%
  \makeatletter%
  \providecommand\color[2][]{%
    \errmessage{(Inkscape) Color is used for the text in Inkscape, but the package 'color.sty' is not loaded}%
    \renewcommand\color[2][]{}%
  }%
  \providecommand\transparent[1]{%
    \errmessage{(Inkscape) Transparency is used (non-zero) for the text in Inkscape, but the package 'transparent.sty' is not loaded}%
    \renewcommand\transparent[1]{}%
  }%
  \providecommand\rotatebox[2]{#2}%
  \ifx\svgwidth\undefined%
    \setlength{\unitlength}{714.35553603bp}%
    \ifx\svgscale\undefined%
      \relax%
    \else%
      \setlength{\unitlength}{\unitlength * \real{\svgscale}}%
    \fi%
  \else%
    \setlength{\unitlength}{\svgwidth}%
  \fi%
  \global\let\svgwidth\undefined%
  \global\let\svgscale\undefined%
  \makeatother%
  \begin{picture}(1,0.51242676)%
    \put(0,0){\includegraphics[width=\unitlength,page=1]{t_of_tau_tex.pdf}}%
    \put(0.0184896,0.4722386){\color[rgb]{0,0,0}\makebox(0,0)[lb]{\smash{$t$}}}%
    \put(0.94271799,0.01889062){\color[rgb]{0,0,0}\makebox(0,0)[lb]{\smash{$\tau '$}}}%
    \put(0.72340032,0.00419329){\color[rgb]{0,0,0}\makebox(0,0)[lb]{\smash{$\dfrac{\hbar\pi}{2mw_{0}\Delta v}$}}}%
  \end{picture}%
\endgroup%

%% file: EGT_tex.pdf_tex
\begingroup%
  \makeatletter%
  \providecommand\color[2][]{%
    \errmessage{(Inkscape) Color is used for the text in Inkscape, but the package 'color.sty' is not loaded}%
    \renewcommand\color[2][]{}%
  }%
  \providecommand\transparent[1]{%
    \errmessage{(Inkscape) Transparency is used (non-zero) for the text in Inkscape, but the package 'transparent.sty' is not loaded}%
    \renewcommand\transparent[1]{}%
  }%
  \providecommand\rotatebox[2]{#2}%
  \ifx\svgwidth\undefined%
    \setlength{\unitlength}{922.39999846bp}%
    \ifx\svgscale\undefined%
      \relax%
    \else%
      \setlength{\unitlength}{\unitlength * \real{\svgscale}}%
    \fi%
  \else%
    \setlength{\unitlength}{\svgwidth}%
  \fi%
  \global\let\svgwidth\undefined%
  \global\let\svgscale\undefined%
  \makeatother%
  \begin{picture}(1,0.63201585)%
    \put(0,0){\includegraphics[width=\unitlength,page=1]{EGT_tex.pdf}}%
    \put(0.00970982,0.60325201){\color[rgb]{0,0,0}\makebox(0,0)[lb]{\smash{$x$}}}%
    \put(0.43590475,0.32194968){\color[rgb]{0,0,0}\makebox(0,0)[lb]{\smash{$t$}}}%
    \put(0.52630329,0.59911226){\color[rgb]{0,0,0}\makebox(0,0)[lb]{\smash{$x'$}}}%
    \put(0.96704525,0.3222439){\color[rgb]{0,0,0}\makebox(0,0)[lb]{\smash{$t$}}}%
    \put(0,0){\includegraphics[width=\unitlength,page=2]{EGT_tex.pdf}}%
    \put(0.01048744,0.29153078){\color[rgb]{0,0,0}\makebox(0,0)[lb]{\smash{$x$}}}%
    \put(0.43668235,0.01022855){\color[rgb]{0,0,0}\makebox(0,0)[lb]{\smash{$t$}}}%
    \put(0.52708087,0.28739103){\color[rgb]{0,0,0}\makebox(0,0)[lb]{\smash{$x'$}}}%
    \put(0.96782288,0.01052272){\color[rgb]{0,0,0}\makebox(0,0)[lb]{\smash{$t$}}}%
    \put(0,0){\includegraphics[width=\unitlength,page=3]{EGT_tex.pdf}}%
  \end{picture}%
\endgroup%

%% file: Linear_potential_tex.pdf_tex
\begingroup%
  \makeatletter%
  \providecommand\color[2][]{%
    \errmessage{(Inkscape) Color is used for the text in Inkscape, but the package 'color.sty' is not loaded}%
    \renewcommand\color[2][]{}%
  }%
  \providecommand\transparent[1]{%
    \errmessage{(Inkscape) Transparency is used (non-zero) for the text in Inkscape, but the package 'transparent.sty' is not loaded}%
    \renewcommand\transparent[1]{}%
  }%
  \providecommand\rotatebox[2]{#2}%
  \ifx\svgwidth\undefined%
    \setlength{\unitlength}{858.60135296bp}%
    \ifx\svgscale\undefined%
      \relax%
    \else%
      \setlength{\unitlength}{\unitlength * \real{\svgscale}}%
    \fi%
  \else%
    \setlength{\unitlength}{\svgwidth}%
  \fi%
  \global\let\svgwidth\undefined%
  \global\let\svgscale\undefined%
  \makeatother%
  \begin{picture}(1,0.31856565)%
    \put(0,0){\includegraphics[width=\unitlength,page=1]{Linear_potential_tex.pdf}}%
    \put(0.66292103,0.28393347){\color[rgb]{0,0,0}\makebox(0,0)[lb]{\smash{$V'(x')$}}}%
    \put(0.96408726,0.12009316){\color[rgb]{0,0,0}\makebox(0,0)[lb]{\smash{$x'$}}}%
    \put(0,0){\includegraphics[width=\unitlength,page=2]{Linear_potential_tex.pdf}}%
    \put(0.15039082,0.28383438){\color[rgb]{0,0,0}\makebox(0,0)[lb]{\smash{$V(x)$}}}%
    \put(0.43367404,0.11999407){\color[rgb]{0,0,0}\makebox(0,0)[lb]{\smash{$x$}}}%
    \put(0,0){\includegraphics[width=\unitlength,page=3]{Linear_potential_tex.pdf}}%
  \end{picture}%
\endgroup%

%% file: Nonlinear_greenberger_transformation_tex.pdf_tex
\begingroup%
  \makeatletter%
  \providecommand\color[2][]{%
    \errmessage{(Inkscape) Color is used for the text in Inkscape, but the package 'color.sty' is not loaded}%
    \renewcommand\color[2][]{}%
  }%
  \providecommand\transparent[1]{%
    \errmessage{(Inkscape) Transparency is used (non-zero) for the text in Inkscape, but the package 'transparent.sty' is not loaded}%
    \renewcommand\transparent[1]{}%
  }%
  \providecommand\rotatebox[2]{#2}%
  \ifx\svgwidth\undefined%
    \setlength{\unitlength}{927.29938177bp}%
    \ifx\svgscale\undefined%
      \relax%
    \else%
      \setlength{\unitlength}{\unitlength * \real{\svgscale}}%
    \fi%
  \else%
    \setlength{\unitlength}{\svgwidth}%
  \fi%
  \global\let\svgwidth\undefined%
  \global\let\svgscale\undefined%
  \makeatother%
  \begin{picture}(1,0.29182289)%
    \put(0,0){\includegraphics[width=\unitlength,page=1]{Nonlinear_greenberger_transformation_tex.pdf}}%
    \put(0.52254814,0.26521582){\color[rgb]{0,0,0}\makebox(0,0)[lb]{\smash{$x$}}}%
    \put(0.93909649,0.10741311){\color[rgb]{0,0,0}\makebox(0,0)[lb]{\smash{$t$}}}%
    \put(0,0){\includegraphics[width=\unitlength,page=2]{Nonlinear_greenberger_transformation_tex.pdf}}%
    \put(0.00752018,0.26592625){\color[rgb]{0,0,0}\makebox(0,0)[lb]{\smash{$x$}}}%
    \put(0.42406853,0.10812354){\color[rgb]{0,0,0}\makebox(0,0)[lb]{\smash{$t$}}}%
    \put(0,0){\includegraphics[width=\unitlength,page=3]{Nonlinear_greenberger_transformation_tex.pdf}}%
  \end{picture}%
\endgroup%

%% file: Quadratic_potentia_tex.pdf_tex
\begingroup%
  \makeatletter%
  \providecommand\color[2][]{%
    \errmessage{(Inkscape) Color is used for the text in Inkscape, but the package 'color.sty' is not loaded}%
    \renewcommand\color[2][]{}%
  }%
  \providecommand\transparent[1]{%
    \errmessage{(Inkscape) Transparency is used (non-zero) for the text in Inkscape, but the package 'transparent.sty' is not loaded}%
    \renewcommand\transparent[1]{}%
  }%
  \providecommand\rotatebox[2]{#2}%
  \ifx\svgwidth\undefined%
    \setlength{\unitlength}{859.94289608bp}%
    \ifx\svgscale\undefined%
      \relax%
    \else%
      \setlength{\unitlength}{\unitlength * \real{\svgscale}}%
    \fi%
  \else%
    \setlength{\unitlength}{\svgwidth}%
  \fi%
  \global\let\svgwidth\undefined%
  \global\let\svgscale\undefined%
  \makeatother%
  \begin{picture}(1,0.31025318)%
    \put(0,0){\includegraphics[width=\unitlength,page=1]{Quadratic_potentia_tex.pdf}}%
    \put(0.66008829,0.28211556){\color[rgb]{0,0,0}\makebox(0,0)[lb]{\smash{$V'(x')$}}}%
    \put(0.96078476,0.11853085){\color[rgb]{0,0,0}\makebox(0,0)[lb]{\smash{$x'$}}}%
    \put(0,0){\includegraphics[width=\unitlength,page=2]{Quadratic_potentia_tex.pdf}}%
    \put(0.14835765,0.28201662){\color[rgb]{0,0,0}\makebox(0,0)[lb]{\smash{$V(x)$}}}%
    \put(0.43119895,0.11843191){\color[rgb]{0,0,0}\makebox(0,0)[lb]{\smash{$x$}}}%
    \put(0,0){\includegraphics[width=\unitlength,page=3]{Quadratic_potentia_tex.pdf}}%
  \end{picture}%
\endgroup%

%% file: moving_walls_schematic2_tex.pdf_tex
\begingroup%
  \makeatletter%
  \providecommand\color[2][]{%
    \errmessage{(Inkscape) Color is used for the text in Inkscape, but the package 'color.sty' is not loaded}%
    \renewcommand\color[2][]{}%
  }%
  \providecommand\transparent[1]{%
    \errmessage{(Inkscape) Transparency is used (non-zero) for the text in Inkscape, but the package 'transparent.sty' is not loaded}%
    \renewcommand\transparent[1]{}%
  }%
  \providecommand\rotatebox[2]{#2}%
  \ifx\svgwidth\undefined%
    \setlength{\unitlength}{650.51389679bp}%
    \ifx\svgscale\undefined%
      \relax%
    \else%
      \setlength{\unitlength}{\unitlength * \real{\svgscale}}%
    \fi%
  \else%
    \setlength{\unitlength}{\svgwidth}%
  \fi%
  \global\let\svgwidth\undefined%
  \global\let\svgscale\undefined%
  \makeatother%
  \begin{picture}(1,0.7068213)%
    \put(0,0){\includegraphics[width=\unitlength,page=1]{moving_walls_schematic2_tex.pdf}}%
    \put(0.30806109,0.66335391){\color[rgb]{0,0,0}\makebox(0,0)[lb]{\smash{$x$}}}%
    \put(0.94756195,0.16129697){\color[rgb]{0,0,0}\makebox(0,0)[lb]{\smash{$t$}}}%
    \put(0.04325514,0.24584712){\color[rgb]{0,0,0}\makebox(0,0)[lb]{\smash{$b(0)$}}}%
    \put(0.21861676,0.39748047){\color[rgb]{0,0,0}\makebox(0,0)[lb]{\smash{$w(0)$}}}%
    \put(0.31514021,0.47892804){\color[rgb]{0,0,0}\makebox(0,0)[lb]{\smash{$w_{2}(t)$}}}%
    \put(0.33904322,0.09250329){\color[rgb]{0,0,0}\makebox(0,0)[lb]{\smash{$w_{1}(t)$}}}%
    \put(0,0){\includegraphics[width=\unitlength,page=2]{moving_walls_schematic2_tex.pdf}}%
    \put(0.77359245,0.3388289){\color[rgb]{0,0,0}\makebox(0,0)[lb]{\smash{$b(t)$}}}%
  \end{picture}%
\endgroup%

%% file: Monomial_well_tex.pdf_tex
\begingroup%
  \makeatletter%
  \providecommand\color[2][]{%
    \errmessage{(Inkscape) Color is used for the text in Inkscape, but the package 'color.sty' is not loaded}%
    \renewcommand\color[2][]{}%
  }%
  \providecommand\transparent[1]{%
    \errmessage{(Inkscape) Transparency is used (non-zero) for the text in Inkscape, but the package 'transparent.sty' is not loaded}%
    \renewcommand\transparent[1]{}%
  }%
  \providecommand\rotatebox[2]{#2}%
  \ifx\svgwidth\undefined%
    \setlength{\unitlength}{796.82851716bp}%
    \ifx\svgscale\undefined%
      \relax%
    \else%
      \setlength{\unitlength}{\unitlength * \real{\svgscale}}%
    \fi%
  \else%
    \setlength{\unitlength}{\svgwidth}%
  \fi%
  \global\let\svgwidth\undefined%
  \global\let\svgscale\undefined%
  \makeatother%
  \begin{picture}(1,0.30581952)%
    \put(0,0){\includegraphics[width=\unitlength,page=1]{Monomial_well_tex.pdf}}%
    \put(0.26271638,0.26827092){\color[rgb]{0,0,0}\makebox(0,0)[lb]{\smash{$w(t)$}}}%
    \put(0.65547317,0.26672418){\color[rgb]{0,0,0}\makebox(0,0)[lb]{\smash{$w(t)$}}}%
    \put(0.46048015,0.01016387){\color[rgb]{0,0,0}\makebox(0,0)[lb]{\smash{$t$}}}%
    \put(0.96844853,0.01219222){\color[rgb]{0,0,0}\makebox(0,0)[lb]{\smash{$t$}}}%
    \put(0.03303296,0.00484354){\color[rgb]{0,0,0}\makebox(0,0)[lb]{\smash{$-T$}}}%
    \put(0.531005,0.00617711){\color[rgb]{0,0,0}\makebox(0,0)[lb]{\smash{$-T$}}}%
  \end{picture}%
\endgroup%

%% file: main.bbl
\begin{thebibliography}{10}

\bibitem{HillWheeler}
David~Lawrence Hill and John~Archibald Wheeler.
\newblock Nuclear constitution and the interpretation of fission phenomena.
\newblock {\em Phys. Rev.}, 89:1102--1145, Mar 1953.

\bibitem{Doescher}
S.~W. Doescher and M.~H. Rice.
\newblock Infinite square-well potential with a moving wall.
\newblock {\em American Journal of Physics}, 37(12):1246--1249, 1969.

\bibitem{Berry:1984}
M~V Berry and G~Klein.
\newblock Newtonian trajectories and quantum waves in expanding force fields.
\newblock {\em Journal of Physics A: Mathematical and General}, 17(9):1805,
  1984.

\bibitem{LevyLeBlond}
Jean-Marc Lévy-Leblond.
\newblock A geometrical quantum phase effect.
\newblock {\em Physics Letters A}, 125(9):441 -- 442, 1987.

\bibitem{Greenberger}
Daniel~M. Greenberger.
\newblock A new non-local effect in quantum mechanics.
\newblock {\em Physica B+C}, 151(1):374 -- 377, 1988.

\bibitem{Pinder}
D.~N. Pinder.
\newblock The contracting square quantum well.
\newblock {\em American Journal of Physics}, 58(1):54--58, 1990.

\bibitem{Makowski:91}
A.J. Makowski and S.T. Dembiński.
\newblock Exactly solvable models with time-dependent boundary conditions.
\newblock {\em Physics Letters A}, 154(5):217 -- 220, 1991.

\bibitem{Pereshogin}
P.~{Pereshogin} and P.~{Pronin}.
\newblock {Effective Hamiltonian and Berry phase in a quantum mechanical system
  with time dependent boundary conditions}.
\newblock {\em Physics Letters A}, 156:12--16, June 1991.

\bibitem{Makowski:92}
A.~J. {Makowski} and P.~{Pep{\l}owski}.
\newblock {On the behaviour of quantum systems with time-dependent boundary
  conditions}.
\newblock {\em Physics Letters A}, 163:143--151, March 1992.

\bibitem{Dodonov}
V.~V. Dodonov, A.~B. Klimov, and D.~E. Nikonov.
\newblock Quantum particle in a box with moving walls.
\newblock {\em Journal of Mathematical Physics}, 34(8):3391--3404, 1993.

\bibitem{MORALES}
Daniel~A. Morales, Zaida Parra, and Rafael Almeida.
\newblock On the solution of the schrödinger equation with time dependent
  boundary conditions.
\newblock {\em Physics Letters A}, 185(3):273 -- 276, 1994.

\bibitem{Cervero}
J.~M. Cerveró and J.~D. Lejarreta.
\newblock The time-dependent canonical formalism: Generalized harmonic
  oscillator and the infinite square well with a moving boundary.
\newblock {\em EPL (Europhysics Letters)}, 45(1):6, 1999.

\bibitem{Campo}
Shortcuts to adiabaticity in a time-dependent box.
\newblock {\em Scientific Reports}, 2(648), 2012.

\bibitem{Berry:1996}
M~V Berry.
\newblock Quantum fractals in boxes.
\newblock {\em Journal of Physics A: Mathematical and General}, 29(20):6617,
  1996.

\bibitem{AronsteinStroud}
David~L. Aronstein and C.~R. Stroud.
\newblock Fractional wave-function revivals in the infinite square well.
\newblock {\em Phys. Rev. A}, 55:4526--4537, Jun 1997.

\bibitem{Berry:2001}
Michael Berry, Irene Marzoli, and Wolfgang Schleich.
\newblock Quantum carpets, carpets of light.
\newblock {\em Physics World}, 14(6):39, 2001.

\bibitem{Niederer}
U.~Niederer.
\newblock {The maximal kinematical invariance group of the free Schrodinger
  equation.}
\newblock {\em Helv. Phys. Acta}, 45:802--810, 1972.

\bibitem{Rosen}
Gerald Rosen.
\newblock Galilean invariance and the general covariance of nonrelativistic
  laws.
\newblock {\em American Journal of Physics}, 40(5):683--687, 1972.

\bibitem{Takagi:1}
Shin Takagi.
\newblock Quantum dynamics and non-inertial frames of reference. i: Generality.
\newblock {\em Progress of Theoretical Physics}, 85(3):463--479, 1991.

\bibitem{BerryPhase}
M.~V. Berry.
\newblock Quantal phase factors accompanying adiabatic changes.
\newblock {\em Proceedings of the Royal Society of London A: Mathematical,
  Physical and Engineering Sciences}, 392(1802):45--57, 1984.

\bibitem{Apostol}
Tom~M. Apostol.
\newblock {\em Introduction to Analytic Number Theory (Undergraduate Texts in
  Mathematics)}.
\newblock Springer, 2010.

\bibitem{Cooney}
Kieran Cooney and Frank~H. Peters.
\newblock Analysis of multimode interferometers.
\newblock {\em Opt. Express}, 24(20):22481--22515, Oct 2016.

\bibitem{Holstein}
Barry~R. Holstein.
\newblock The extended galilean transformation and the path integral.
\newblock {\em American Journal of Physics}, 51(11):1015--1016, 1983.

\bibitem{Klink}
W.H. Klink.
\newblock Quantum mechanics in nonintertial reference frames.
\newblock {\em Annals of Physics}, 260(1):27 -- 49, 1997.

\bibitem{MacGregor}
B.~R. MacGregor, A.~E. McCoy, and S.~Wickramasekara.
\newblock Unitary representations of the galilean line group: Quantum
  mechanical principle of equivalence.
\newblock 2011.

\bibitem{Takagi:2}
Shin Takagi.
\newblock Quantum dynamics and non-inertial frames of references. ii: Harmonic
  oscillators.
\newblock {\em Progress of Theoretical Physics}, 85(4):723--742, 1991.

\bibitem{Takagi:3}
Sin Takagi.
\newblock Quantum dynamics and non-inertial frames of reference. iii: Charged
  particle in time-dependent uniform electromagnetic field.
\newblock {\em Progress of Theoretical Physics}, 86(4):783--798, 1991.

\bibitem{Bender}
Carl Bender.
\newblock {\em Advanced Mathematical Methods for Scientists and Engineers I :
  Asymptotic Methods and Perturbation Theory}.
\newblock Springer New York, New York, NY, 1999.

\bibitem{Appell}
Appell M.P.
\newblock Sur l'\'{e}quation $\frac{\partial ^2z}{\partial x^2}-\frac{\partial
  z}{\partial y}=0$ et la th\'{e}orie de la chaleur.
\newblock {\em J. Math. Pure Appl.}, 8:187--216, 1892.

\bibitem{Widder}
D.V. Widder.
\newblock {\em The Heat Equation}.
\newblock Pure and Applied Mathematics. Elsevier Science, 1976.

\bibitem{Shankar}
Ramamurti Shankar.
\newblock {\em Principles of Quantum Mechanics}.
\newblock Springer New York, Boston, MA, 1994.

\end{thebibliography}
